\documentclass{aa} 
\usepackage{psfig}

\newcommand{\bm}[1]{\mbox{{\boldmath $#1$}}}

\newcommand{\pder}[3]{\frac{{\partial}^{#3} {#1}}{{\partial} {#2}^{#3}}}

\begin{document}

\title{Stability of hydrodynamical relativistic planar jets. \\ 
  I. Linear evolution and saturation of Kelvin-Helmholtz modes.}

\subtitle{}

\author{M. Perucho\inst{1}
          \and
	M. Hanasz\inst{2}
          \and 
	J. M. Mart\'{\i}\inst{1}
          \and
        H. Sol\inst{3}
 }

   \offprints{M. Perucho}

   \institute{Departamento de Astronom\'{\i}a y Astrof\'{\i}sica,
              Universidad de Valencia, 46100 Burjassot (SPAIN)\\
              \email{manuel.perucho@uv.es, jose-maria.marti@uv.es}
              \and Toru\'n Centre for Astronomy, Nicholas Copernicus
              University, PL-97-148 Piwnice k. Torunia (POLAND)\\
              \email{mhanasz@astri.uni.torun.pl} \and LUTH,
              Observatoire de Paris-Meudon, Pl. J. Jansen 5,Meudon
              92195 (FRANCE)\\ \email{Helene.Sol@obspm.fr} }

\date{Received .../ Accepted ...}

\abstract{The effects of relativistic dynamics and thermodynamics in
the development of Kelvin-Helmholtz instabilities in planar,
relativistic jets along the early phases (namely linear and saturation
phases) of evolution has been studied by a combination of linear
stability analysis and high-resolution numerical simulations for the
most unstable first reflection modes in the temporal approach. Three
different values of the jet Lorentz factor (5, 10 and 20) and a few
different values of specific internal energy of the jet matter (from
$0.08$ to $60.0 c^2$) have been considered. Figures
illustrating the evolution of the perturbations are also shown.

  Our simulations reproduce the linear regime of evolution of the
excited eigenmodes of the different models with a high accuracy.  In
all the cases the longitudinal velocity perturbation is the first
quantity that departs from the linear growth when it reaches a value
close to the speed of light in the jet reference frame. The saturation
phase extends from the end of the linear phase up to the saturation of
the transversal velocity perturbation (at approximately $0.5c$ in the
jet reference frame). The saturation times for the different numerical
models are explained from elementary considerations, i.e. from
properties of linear modes provided by the linear stability analysis
and from the limitation of the transversal perturbation velocity.  The 
limitation of the components of the velocity
perturbation at the end of the linear and saturation phases allows us
to conclude that the relativistic nature of the flow appears to be
responsible for the departure of the system from linear evolution.

  The high accuracy of our simulations in describing the
early stages of evolution of the KH instability (as derived from the
agreement between the computed and expected linear growth rates and
the consistency of the saturation times) establishes a solid basis to
study the fully nonlinear regime, to be done elsewhere. The present
paper also sets the theoretical and numerical background for these
further studies.

\keywords{Galaxies: jets - relativistic hydrodynamics - instabilities} }

\authorrunning{M. Perucho et al.}
\titlerunning{Stability of hydrodynamical relativistic planar jets. I}

\maketitle

\section{Introduction}

  Relativistic jets are found in many astrophysical objects like
active galactic nuclei (see eg. Zensus 1997), microquasars (Mirabel \&
Rodriguez 1999) and gamma-ray bursts (Sari, Piran \& Halpern
1999). The basic physical property of jets is their huge flux of
kinetic energy, taken away from compact unresolved or marginally
resolved central regions.  The presence of jets in such objects
provides a unique opportunity to study the physics of these central
regions through investigations of the jet flows.  However, the
extraction of information about the central object is not a simple
task, because jets interact in complex ways with the surrounding
interstellar and intergalactic medium. On one hand the dynamical
interaction of the jet matter with the ambient medium leads to the
formation of shocks, turbulence, acceleration of charged particles and
subsequent emission of a broad-range electromagnetic radiation, which
makes jets observable. On the other hand the complex nature of the
interaction which is due to the presence of flow instabilities,
especially the Kelvin-Helmholtz (KH) instability, makes it difficult
to distinguish those properties which are directly related to
the central object from those which are due to the interaction of jets
with the ambient medium. As an example one can invoke the wavelike
patterns in jets, which may result from the precession of the rotation
axis of the accretion disk at the jet base, or from KH instability,
which finds favorable conditions at the interface of jet and external
medium (Trussoni, Ferrari \& Zaninetti, 1983).  Very recently, the KH
instability theory has been successfully used to interpret the
structure of the pc-scale jet in the radio source 3C273 (Lobanov \&
Zensus 2001).

  At kiloparsec scales, the surprisingly stable propagation of
relativistic jets in some sources (e.g., Cyg~A) contrasts with the
deceleration and decollimation observed in other sources (e.g. 3C31).
These two sources are representative examples of two kinds, which are
classified into Fanaroff-Riley type I (FRI) and Fanaroff-Riley type II
(FRII) radio sources (Fanaroff \& Riley 1974). The
morphological dichotomy of FRI and FRII sources may be related to the
stability properties of relativistic jets with different kinetic
powers.

  This complex situation motivated us to study the interaction of
relativistic jets with their ambient medium and more specifically the
KH instability in detail, by applying a linear stability analysis
along with numerical simulations.

  The linear analysis of KH instability in relativistic jets started
with the work of Turland \& Scheuer (1976) and Blandford \& Pringle
(1976) who derived and solved a dispersion relation for a single plane
boundary between the relativistic flow and the ambient
medium. Next, Ferrari, Trussoni \& Zaninetti (1978) and Hardee (1979)
examined properties of KH instability in relativistic cylindrical jets
by following the derivation of the dispersion relation in the
nonrelativistic case, in the vortex sheet approximation, done by Gill
(1965). They numerically solved the dispersion relation, found
unstable KH modes and classified them into the fundamental (surface) and
reflection (body) family of modes.  The classification is related to
the number of nodes, across the jet, of sound waves reflecting in
between jet boundaries. An internal wave pattern is formed by the
composition of oblique waves, for which the jet interior is a resonant
cavity. The physical meaning of KH instability in supersonic jets has
been discussed by Payne \& Cohn (1985), who have shown that the
presence of instability is associated with the overreflection of
soundwaves (the modulus of reflection coefficient is larger than 1) on
the sheared jet boundaries.  

  Subsequent studies include effects of magnetic field (Ferrari,
Trussoni \& Zaninetti 1981), the effects of the shear layer, replacing the
vortex sheet in the nonrelativistic planar case (Ferrari, Massaglia \&
Trussoni 1981; Ray 1982; Choudhury \& Lovelace 1984), and conical and
cylindrical jet geometry (Birkinshaw 1984; Hardee 1984, 1986,
1987). The effects of cylindrical a shear layer have been examined by
Birkinshaw (1991) in the nonrelativistic case, attempted by Urpin
(2002) in the relativistic case and the presence of multiple
components (jet + sheath + ambient medium) was investigated by Hanasz
\& Sol (1996, 1998).  Other authors have investigated current-driven modes
in magnetized jets (Appl \& Camenzind 1992; Appl 1996) in addition to
KH modes.

  An extension of the linear stability analysis in the relativistic
case to the weakly nonlinear regime has been performed by Hanasz
(1995) and led to the conclusion that Kelvin-Helmholtz instability
saturates at finite amplitudes due to various nonlinear effects. An
explanation of the nature of the mentioned nonlinearities has been
proposed by Hanasz (1997). The most significant effect results from
the relativistic character of the jet flow, namely from the fact that
the velocity perturbation cannot exceed the speed of light.

  A more recent approach is to perform a linear stability analysis in
parallel with numerical simulations and to compare the results of
both methods in the linear regime and then to follow the nonlinear
evolution of the KH instability resulting from numerical
simulations. Hardee \& Norman (1988) and Norman \& Hardee (1988) have
made such a study for nonrelativistic jets in the spatial approach and
Bodo et al. (1994) in the temporal approach. In the relativistic case
this kind of approach was applied for the first time by Hardee et
al. (1998) in the case of axisymmetric, cylindrical jets and then
extended to the 3D case by Hardee et al. (2001).

  Similarly to the linear stability analysis, the numerical
simulations of jet evolution can be performed following both the
spatial and the temporal approach, depending on the particular choice
of initial and boundary conditions. In the temporal approach one
considers a short slice of jet limited by periodic boundaries along
the jet axis, and adds some specific perturbation, eg. an eigenmode
resulting from the linear stability analysis. Due to the periodic
boundary conditions the growing perturbations can only be composed of
modes having a wavelength equal to the length of the computational
box and/or its integer fractions (Bodo et al. 1994, 1995, 1998).
Whereas the spatial approach appears more appropriate to analyze the
global dynamics and morphology of the whole jet, the temporal approach
is suitable for the comparison between the numerical results and
analytical studies of the jet stability because, due to the fact that
only part of the jet is simulated, a high effective numerical
resolution is achievable with limited computer resources.

  Numerical simulations (Mart\'{\i} et al. 1997; Hardee et al. 1998,
Rosen et al.  1999) demonstrate that jets with high Lorentz factors
and high internal energy are influenced very weakly by the
Kelvin-Helmholtz instability.  Moreover, Hardee et al. (1998), Rosen
et al. (1999) note that contrary to the cases with lower Lorentz
factors and lower internal energies, the relativistically hot and high
Lorentz factor jets do not develop modes of KH instability predicted
by the linear theory.  They interpret this fact as the result of a
lack of appropriate perturbations generating the instability in the
system. In the limit of high internal
energies of the jet matter the Kelvin-Helmholtz instability is
expected to develop with the highest growth rate.

  In this paper we focus on the role of internal energy and the Lorentz
factor for the stability of relativistic planar two-dimensional jets
in the temporal approach. We chose this simple configuration in order
to enhance our ability to understand the complex nonlinear evolution
of the unstable KH modes in relativistic jets.  For this purpose, we
perform a linear analysis of the KH instability and construct
perturbations analytically: the eigenmodes of the linear problem
representing the most unstable first reflection mode for each set of
jet parameters. These modes are then incorporated as initial states
of numerical simulations.
The choice of planar two-dimensional jets allows us
to study symmetric and antisymmetric modes of KH instability which
resemble, respectively, pinch and helical modes of cylindrical
jets. Only the flutting modes of cylindrical jets, corresponding to
azimuthal wavenumbers larger than 1, have no counterparts among the
eigenmodes of planar two-dimensional jets. In the present paper we
investigate only the symmetric modes.

  Our aim is to investigate physical details of the process of
transition of KH instability modes from the linear to the nonlinear
regime. In the next step (Paper II) we shall investigate the relations
of long-term nonlinear evolution of KH instability (including the
mixing phase and the properties of the evolved flow at the end of our
simulations) on the initial jet parameters.

  The paper is organized as follows. In Section~\ref{sect:method} we
describe our method, i.e. perform the linear analysis of the KH
instability for the relativistic planar case, then define
perturbations and describe numerical algorithm and other
details of numerical simulations. In Section~\ref{sect:results} we
present and discuss results of the both linear stability analysis and
the numerical simulations. In Section~\ref{s:concl} we
summarize and conclude our paper.

\section{Method \label{sect:method}}

  Following the standard procedure (see eg. Gill 1965; Ferrari, Trussoni \&
Zaninetti 1978; Hardee 1979) we derive the dispersion relation for the
Kelvin-Helmholtz modes. We focus on the simplest geometrical configuration of
two-dimensional planar relativistic jets and apply the temporal stability
analysis. 

\subsection{Linear stability analysis}

  The full set of equations describing the current problem consists of
the set of relativistic equations of hydrodynamics for a perfect fluid, (e.g.
Ferrari, Trussoni \& Zaninetti 1978)
\begin{equation}
\gamma^2 \!\!\left( \rho \!+\! {p\over c^2} \right)\!\! \left[\!
{\partial \vec v \over \partial t}\!+\! (\vec v  \cdotp \! \vec\nabla)
\vec v \! \right]
\! + \vec\nabla p + {\vec v \over c^2} {\partial p \over \partial t}
\! = \! 0, \label{eulereq}
\end{equation}
\begin{equation}
\gamma \!\!\left( {\partial \rho\over \partial t} \!+\! \vec v \cdotp
\vec \nabla\rho \right) \!+\! \left( \rho + {p\over c^2} \right)\!\!
\left[ {\partial \gamma \over \partial t} + \vec\nabla \cdotp (\gamma \vec v)
\right] \!\!= \! 0, \label{conteq}
\end{equation}
\noindent
and an (adiabatic) equation of state 
\begin{equation}  
p \rho_0^{-\Gamma} = \mbox{{\rm const}}. \label{eqos}
\label{eos}
\end{equation}
In the preceding equations, $c$ is the speed of light, $\rho_0$ is the
particle rest mass density (i.e., $\rho_0 \equiv m n$, where $m$ is
the particle rest mass and $n$ the number density in the fluid rest
frame). $\rho$ stands for the relativistic density which is related to
the particle rest mass density and the specific internal energy,
$\varepsilon$, by $\rho = \rho_0(1+\varepsilon/c^2)$. The enthalpy is
defined as $w = \rho + p/c^2$, the sound speed is given by $c_s^2 =
\Gamma p/w$, where $\Gamma$ is the adiabatic index. The relation
between pressure and the specific internal energy is $p =(\Gamma - 1)
\varepsilon \rho_0$. The velocity of the fluid is represented by $\vec
v$ and $\gamma$ is the corresponding Lorentz factor.

\begin{figure*} 
\centerline{
\psfig{file=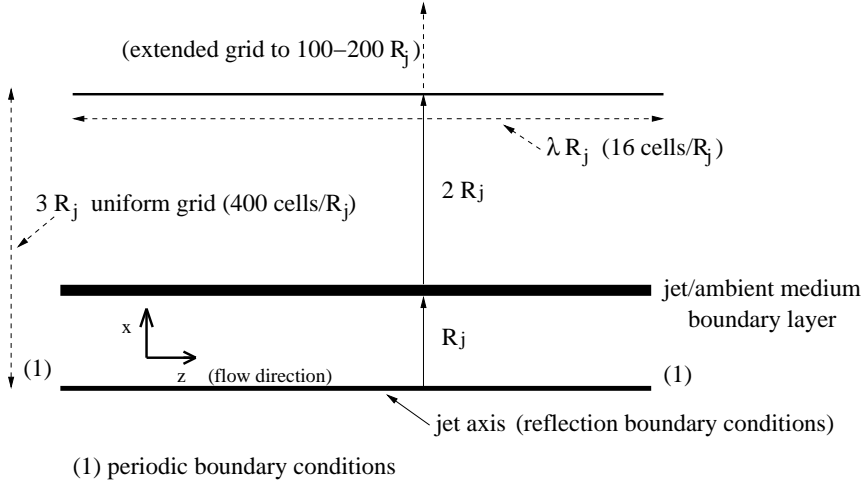,width=0.8\textwidth,angle=270,clip=0} 
}
\caption{The geometry of the flow considered in the linear
stability analysis and the numerical simulations, including a
description of the boundary conditions.} 

\label{f:sketch}

\end{figure*}

  The assumed geometry of the jet considered in the forthcoming linear
stability analysis and the numerical simulations is sketched in
Fig.~\ref{f:sketch}.
First of all the considered jet is 2D-planar and
symmetric with respect to the $x=0$ plane. The flow in the jet moves
in the positive $z$ direction and its matter forms a contact
discontinuity (a vortex sheet) with the matter of external medium at
$x= -R_{\rm j}$ and $x=R_{\rm j}$. From now on all quantities
representing the jet will be assigned with the '$j$' subscript and the
quantities representing the ambient medium will be assigned with
'$a$'.

%
%

  The following matching conditions are imposed on the interface
between the jet and the ambient medium
\begin{eqnarray}
p_a = p_j   &  \mbox{for} &  |x|=R_j \label{pmatch} \\ 
h_a = h_j   &  \mbox{for} &  |x|=R_j.\label{hmatch}
\end{eqnarray}
The matching conditions express the assumption of equality of the jet and
ambient pressures and the equality of transversal displacements of jet
($h_j$) and ambient ($h_a$) fluid elements adjacent to the jet
boundary (at $|x| = R_j$).

  In addition, the Sommerfeld radiation condition (expressing
the disappearance of perturbations at the infinity) will be applied for
linear perturbations.

\subsubsection{Equilibrium state for the linear stability analysis
\label{sect:equilibrium}}

  We assume that the initial state is an equilibrium
configuration. The initial equilibrium can be described by the
following set of independent parameters: the Lorentz factor
corresponding to the unperturbed longitudinal jet velocity, $v_j$,
$\gamma = (1 - v_j^2/c^2)^{-1/2}$, the particle rest mass density
of the jet $\rho_{0j}$ (the particle rest mass density of the ambient
medium is normalized to unity: $\rho_{0a} = 1$) and the specific
internal energy of the jet $\varepsilon_j$. The ambient medium is
assumed to be at rest ($v_a = 0$).

  The other dependent parameters describing the equilibrium state are:
the internal jet Mach number $M_j = {v_j}/{c_{sj}}$ corresponding
to the initial jet longitudinal velocity, the relativistic rest mass
density contrast $\displaystyle{\nu =
\frac{\rho_{0j}(1+\varepsilon_j/c^2)}
{\rho_{0a}(1+\varepsilon_a/c^2)}}$ (or, equivalently, the enthalpy
contrast $\displaystyle{\eta=
\frac{\rho_{0j}\left(1+(\Gamma_j-1)\varepsilon_j/c^2/
(1+\varepsilon_j/c^2)\right)}
{\rho_{0a}\left(1+(\Gamma_a-1)\varepsilon_a/c^2/
(1+\varepsilon_a/c^2)\right)}}$) and the specific internal energy of
the ambient medium $\displaystyle{\varepsilon_a = \frac{(\Gamma_j - 1)
\rho_{0j}}{(\Gamma_a - 1) \rho_{0a}} \varepsilon_j}$, which is related
to the specific internal energy of the jet through the pressure
balance condition.

\subsubsection{Dispersion relation}

  The first step towards the dispersion relation is to reduce the
equations (\ref{eulereq}) - (\ref{hmatch}) to the dimensionless form
through the normalization of spatial coordinates to the jet radius
$R_j$, velocities to the sound speed of the jet material $c_{sj}$,
time to the dynamical time $R_j/c_{sj}$ and pressure to the
equilibrium pressure.

  The next step is to decompose each dependent quantity into the
equilibrium value and the linear perturbation. After the reduction of
equations to the dimensionless form and substitution of the perturbed
quantities in equations (\ref{eulereq})-(\ref{hmatch}), we obtain the
following set of dimensionless linearized equations. In the following
the dimensionless quantities will be assigned the same symbols as
the previous dimensional ones. The subscript '$1$' stands for the
linear perturbation of the corresponding variable. For the jet medium
we get

\begin{equation}\label{linjet1}
\Gamma_j \, \gamma^2 \left(\frac{\partial{\bm{v}_{j1}}}{\partial t} +
(\bm{v}_j \cdot \nabla) \bm{v}_{j1} \right) + \nabla p_{j1} +
\frac{\bm{v}_j}{c^2} \frac{\partial p_{j1}}{\partial t} = 0,
\end{equation}

\begin{eqnarray}\label{linjet2}
\frac{\partial p_{j1}}{\partial t} + \bm{v}_j \cdot \nabla p_{j1} +
\Gamma_j \nabla \cdot \bm{v}_{j1} \hspace{4cm} \\ + \gamma^2 \,
\Gamma_j \frac{\bm{v}_j}{c^2} \left(\frac{\partial
\bm{v}_{j1}}{\partial t} + (\bm{v}_j \cdot \nabla
\bm{v}_{j1})\right) = 0 \quad ,\nonumber
\end{eqnarray}

\noindent
where $\bm{v}_j$ is the initial (unperturbed) jet velocity in units
of jet internal sound speed, $c$ is the speed of light in units of the
sound speed and the normalized pressure $p_0=1$ is omitted in the
equations. For the ambient medium we get

\begin{equation}\label{linam1}
\frac{\Gamma_j}{\eta} \frac{\partial
{\bm{v}_{a1}}}{\partial t}  + \nabla p_{a1} = 0,
\end{equation}

\begin{equation}\label{linam2}
\frac{\partial p_{a1}}{\partial t} 
+ \Gamma_a\nabla \cdot \bm{v}_{a1} = 0.
\end{equation}

\noindent
The linearized matching conditions (\ref{pmatch}) and (\ref{hmatch})
at the jet interface read

\begin{eqnarray}
p_{a1} = p_{j1}  &  \mbox{for} &  |x|=1 \label{pmatch1} \\ 
h_{a1} = h_{j1}   &  \mbox{for} &  |x|=1 \label{hmatch1},
\end{eqnarray}
where the displacements of fluid elements adjacent to the contact
interface in the linear regime are related to transversal velocities
by the following formulae
\begin{equation}
v_{jx1} = \left( \pder{}{t}{} + v_j \pder{}{z}{}   \right) h_{j1},
\label{hj} 
\end{equation}
\begin{equation}
v_{ax1} =\pder{}{t}{} h_{a1}. 
\label{ha}
\end{equation}

\noindent 
The following wave equations can be derived from the equations
(\ref{linjet1}) - (\ref{linam2}), respectively for the jet

\begin{eqnarray} \label{pj}
\gamma^2 \left(\pder{}{t}{}+v_j\pder{}{z}{} \right)^2 p_{j1} -
\pder{p_{j1}}{x}{2} \hspace{3cm}\\ - \gamma^2\left( \pder{}{z}{} +
\frac{v_j}{c^2} \pder{}{t}{}\right)^2 p_{j1} =0, \nonumber
\end{eqnarray}

\noindent
and for the ambient medium

\begin{equation} \label{pa}
\pder{p_{a1}}{t}{2} - \frac{\eta\Gamma_a}{\Gamma_j}\left(
\pder{p_{a1}}{x}{2}+ \pder{p_{a1}}{z}{2}\right) = 0.
\end{equation}
It is apparent that Eqn.~(\ref{pj}) and~(\ref{pa}) describe
propagation of oblique sound waves in the jet and ambient medium
respectively.

\noindent
The next step is to substitute perturbations of the form
\begin{equation} \label{eq:deltab}
\delta_{j1} = [ \delta^+_{j} F^+_{j} (x) + \delta^-_{j} F^-_{j} (x) ]
    \exp i (k_{\parallel } z - \omega t) + c.c.
\end{equation}
with $F^{\pm}_{j} (x) = \exp (\pm i\, k_{{j\perp}} x)$ to describe
waves propagating in positive and negative $x$-directions in the beam.
Here we use $k_{\parallel}$ and $k_{a,j\perp}$ for longitudinal and
transverse wavenumbers. Perturbations in the external medium are of
the form
\begin{equation} \label{eq:deltaa}
\delta_{a1} = \delta^+_a F^+_a (x) \exp\, i (k_{\parallel } z-\omega
t) +c.c.
\end{equation}
where $F^+_a (x) = \exp (i\, k_{a\perp} x)$ for $|x|>1$. Assuming that
${\rm Re} (k_{a\perp}) > 0$, only outgoing waves are present in the
ambient medium.  After the substitution of the explicit forms of
pressure perturbations to equations (\ref{pj}) and (\ref{pa}) we
obtain the following expressions for the perpendicular components of
wavevectors

\begin{equation}
k_{a \perp} = \left(\frac{\Gamma_j}{\eta \Gamma_a} \omega^2 -
k_\parallel^2\right)^{1/2}
\end{equation}

\begin{equation}  
k_{j \perp} = ({\omega'}^2-{k'_\parallel}^2)^{1/2}  
\end{equation}
which are standard relations for linear sound waves in both media
(note that $\eta \Gamma_a/\Gamma_j$ is the squared sound speed of
ambient medium in units of the sound speed of jet), $\omega' =
\gamma(\omega - v_j k_\parallel)$ and $k'_{\parallel} =
\gamma(k_\parallel - \frac{v_j}{c^2} \omega)$ are frequency and
wavenumber of the internal sound wave in the reference frame comoving
with the jet. 
The wavevectors $\bm{k}_j = (k_{j\perp}, k_{j\parallel})$ and
$\bm{k}_a = (k_{a\perp}, k_{a\parallel})$ determine the direction of
propagation of sound waves in the jet and ambient medium
respectively. Vanishing of $k_{j\perp}$, for instance, would mean that
the jet internal sound waves move in the axial direction. In cases
$k_{j\perp} \ne 0 $ and/or $k_{a\perp} \ne 0 $ the propagation of the
sound waves is oblique with respect to the jet axis.

  The corresponding perturbation of the jet surface can be written as
\begin{equation}
h_{a1}=h_{j1}= h \exp\, i (k_{\parallel  } z-\omega t) +c.c. ,
\end{equation}
which after the substitution to equations (\ref{hj}) and (\ref{ha})
reads
\begin{equation}
v_{jx1} = -i \left(\omega - v_j k_{\parallel}\right) h_{j1} ,
\end{equation}
\begin{equation}
v_{ax1} = -i \omega  h_{j1}.
\end{equation}

  With the aid of the above perturbations the whole system of partial
differential equations is reduced to a set of homogeneous linear
algebraic equations. The dispersion relation appears as solvability
condition of the mentioned set of equations, namely it arises from
equating the determinant of the linear problem to zero. Within the
present setup, the dispersion relation for the Kelvin-Helmholtz
instability in supersonic, relativistic, two-dimensional slab jets can
be written as
\begin{eqnarray}
\frac{1}{\nu \Gamma_j} \frac{\omega}{\omega'} \frac{({\omega'}^2 -
{k'}^2_{\parallel})^{1/2}} {\left(\frac{\omega^2}{\nu \Gamma_a} -
k_\parallel^2\right)^{1/2}} =
  \hspace{4cm}\\
  \hspace{2cm} =\left\{ 
\begin{array}{lll}
 \!\!{\rm coth}\ i ({\omega'}^2 - {k'}^2_{\parallel})^{1/2} & {\rm
   for} & s = 1\\ & {\rm }& \\ \!\!{\rm th}\ i({\omega'}^2 -
   {k'}^2_{\parallel})^{1/2} & {\rm for}& s=-1\\
\end{array} \nonumber
  \right. \label{disprel}
\end{eqnarray}
where $s = \pm 1$ is the symmetry of perturbation. In the present
paper we focus on symmetric perturbations ($s=1$).

  The above dispersion relation is solved numerically with the aid of
the Newton-Raphson method (see, e.g., Press et al. 1992).  In this way
we obtain the complex frequency as a function of the parallel
component of the wavenumber $k_\parallel$, since we choose the
temporal stability analysis for the present investigations.

\subsubsection{Eigenstates of the system \label{sect:eigenstates}}

  Once the solutions of the dispersion relation $\omega(k)$ are found,
the derivation of eigenstates of the system is straightforward. By
utilizing the set of relativistic equations (\ref{linjet1}) -
(\ref{linam2}) one can relate perturbations of gas density and
velocity to the perturbation of pressure.

  First, we choose the amplitude of the pressure wave in the jet,
$p_j^+$, as a free constant and then relate the other constants to its
value. The corresponding '$-$' amplitudes are computed by multiplying
these values by $s$ in case of pressure and longitudinal velocity and
by $-s$ in the case of transversal velocity.

  The amplitude of pressure wave in the ambient medium, $p_a^+$, is
found from the pressure matching condition at $x=1$:

\begin{equation}
p_{a}^+\exp(i\,k_{a\perp}) = p_{j}^+ \exp(i\,k_{j\perp})+s\,p_{j}^- 
\exp(-i\,k_{j\perp}).
\end{equation}

\noindent
The remaining amplitudes of velocity perturbations are

\begin{equation}
v_{ax}^{+}= \frac{\eta}{\Gamma_j}
\frac{k_{a\perp}}{\omega} p_a^+, 
\end{equation}

\begin{equation}
v_{jx}^{+}= \frac{k_{j\perp}}{\Gamma_j \gamma
\omega'} p_j^{+}, \label{pj-vjperp}
\end{equation}

\begin{equation}
v_{az}^{+} = \frac{\eta}{\Gamma_j}
\frac{k_{\parallel}}{\omega} p_a^+, 
\end{equation}

\begin{equation}
v_{jz}^{+} = \frac{k'_{\parallel}}{\Gamma_j \gamma^2
\omega'} p_j^+,
\end{equation}
where $\eta$ is the ratio of enthalpies as defined earlier.
  
  Regarding specific internal energy and rest-mass density,
perturbations are calculated as follows:

\begin{equation}
\frac{\varepsilon_{a,j1}}{\varepsilon_{a,j}}= \frac{\Gamma_{a,j}-
1}{\Gamma_{a,j}} p_{a,j1}
\end{equation}


\begin{equation}
\frac{\rho_{a,j1}}{\rho_{a,j}+1/c^2}=\frac{1}{\Gamma_{a,j}} p_{a,j1}, 
\end{equation}
where $\rho_{a,j}$ refer to the relativistic density in ambient and
jet media in dimensionless units, respectively, and $\rho_{a,j1}$
is the corresponding perturbation.

\subsubsection{Lorentz transformation of velocity perturbations}

  In the forthcoming discussion of results of numerical simulations we
shall analyze the velocity components in the reference frame comoving
with the jet.  The standard Lorentz transformation formulae for the
velocity components are
\begin{eqnarray} 
v'_z = \frac{v_z - v_j}{1- v_z v_j/c^2}, && 
v'_x = \frac{{v_x}\sqrt{1-v_j^2/c^2}}{1-v_z v_j/c^2}. 
\label{eq:lorentz}
\end{eqnarray} 
Assuming that the Lorentz transformation is applied for linear
perturbations of sufficiently small amplitude, these transformation
rules can be approximated by the following formulae
\begin{eqnarray} 
v'_{z1} = \gamma^2 v_{z1}, && 
v'_{x1} = \gamma   v_{x1},  \label{lorentz-pert}
\end{eqnarray} 
where $v_{x1}$, $v_{z1}$ are the velocity perturbations in the
reference frame of the ambient medium and $v'_{x1}$, $v'_{z1}$ are the
velocity perturbations in the jet reference frame. These
transformation rules tell us that the longitudinal and transversal
velocity perturbations in the jet reference frame are larger by
factors of $\gamma^2$ and $\gamma$ respectively than the corresponding
components in the rest (ambient medium) frame.

\subsection{Numerical simulations} 
\label{ss:numsim}

  Now we proceed by performing numerical simulations of jet models
to study the growth of unstable modes through the
linear and non-linear phases. To this aim, we adopt as an equilibrium
initial state the one described in the former section. In order to
avoid the growth of random perturbations with wavelengths of the order
of the cell size, we replace the transverse discontinuous profiles of
equilibrium quantities by smooth profiles of the form

\begin{equation}
v_z(x) = \frac{v_j}{\cosh(x^m)} \label{shearlayer1}
\end{equation}

\begin{equation}
\rho_0(x) = \rho_{0a} - \frac{\rho_{0a} - \rho_{0j}}{\cosh(x^m)} 
\label{shearlayer2}
\end{equation}
(see Bodo et al. 1994), where $m$ is the steepness parameter for the
shear layer. Typically we used $m=40$, for which the shear layer has a width 
$\sim 0.1 R_j$ (see Appendix A).

On the other hand the linear solutions derived in
Section~\ref{sect:eigenstates} correspond to an idealized case with a
contact discontinuity at the jet interface. Finding of analogous
solutions for the corresponding sheared boundary problem is more
difficult from the numerical point of view (see Ray 1982, and
Choudhury \& Lovelace 1984), therefore we decided to implement
corresponding smoothed eigenstates obtained for the vortex sheet limit
as approximate solutions for the case of sheared boundary.  The
validity of this procedure is proven {\em a posteriori} by the
convergence of theoretical growth rates and the ones determined for
simulated KH modes. Since, in case of the vortex sheet approximation,
all first order dynamical variables (except pressure and the
transversal displacement of fluid element adjacent to the jet
boundary) are discontinuous, we apply the following smoothing of the
equilibrium quantities and linear perturbations at the thin layer
surrounding the jet boundary

\begin{equation}
\delta_1=\delta_{a1}-\frac{\delta_{a1}-\delta_{j1}}{\cosh(x^m)},
\label{smooth-pert}
\end{equation} 

\noindent 
where $\delta$ stands for any perturbed variable. Such smoothed
eigenstates are subsequently implemented as small-amplitude
perturbations in the initial equilibrium model of the simulations.

  Our first aim is to reproduce the linear evolution of unstable modes
by means of hydrodynamical simulations.  For this reason and in order
to make the picture of evolution of the KH instability as clear and
simple as possible, we perturb the equilibrium with the most unstable
first reflection mode, which is smoothed at the jet interface
according to formula (\ref{smooth-pert}). The length of the
computational grid is in each simulation equal to the wavelength of
the mentioned mode. Defining the initial state in such a way, we know
precisely which growth rate to expect in numerical simulations. The
difference between the expected and computed growth rates is a measure
of the performance of the code in describing the linear regime and
gives us confidence on the numerical results from the non-linear
evolution.

  The simulations have been performed using a high--resolution shock
capturing code to solve the equations of relativistic hydrodynamics in
Cartesian planar coordinates. The code is an upgrade of that used to
study large scale (Mart\'{\i} et al. 1997) as well as compact
relativistic jets (G\'omez et al.  1997). The equations of
relativistic hydrodynamics are written in conservation form and the
time variation of the proper rest mass, momentum and total energy
within numerical cells are calculated through the fluxes across the
corresponding cell interfaces. Fluxes are calculated with an
approximate Riemann solver that uses the complete characteristic
information contained in the Riemann problems between adjacent cells
(Donat \& Marquina 1996). It is based on the spectral decomposition of
the Jacobian matrices of the relativistic system of equations derived
in Font et al. (1994) and uses analytical expressions for the left
eigenvectors (Donat et al. 1998). The spatial accuracy of the
algorithm is improved up to third order by means of a conservative
monotonic parabolic reconstruction of the pressure, proper rest-mass
density and the spatial components of the fluid four-velocity, as in
{\tt GENESIS} (see Aloy et al. 1999, and references
therein). Integration in time is done simultaneously in both spatial
directions using a total-variation-diminishing (TVD) Runge-Kutta
scheme of high order (Shu \& Osher 1988). See Mart\'{\i} et al. (1997)
for the differential equations as well as their finite--difference
form, and a description of exhaustive testing of the hydrodynamical
code. Besides relativistic density, momentum and energy, the code also
evolves a passive scalar representing the jet mass fraction. This
allows us to distinguish between ambient and jet matter helping us to
characterize processes like jet/ambient mixing or momentum exchange.

  The parameters of the simulations presented in this paper are listed
in Table~\ref{tab:param}. Values of the parameters were chosen to be
close to those used in some simulations by Hardee et al. (1998) and
Rosen et al. (1999) and are chosen to span a wide range in
thermodynamical properties as well as beam flow Lorentz factors. In
all simulations, the density in the jet and ambient gases are
$\rho_{0j}=0.1$, $\rho_{0a}=1$ respectively and the adiabatic exponent
$\Gamma_{j,a}=4/3$.

  Since the internal rest mass density is fixed, there are two free
parameters characterizing the jet equilibrium: the Lorentz factor and 
the jet specific internal energy displayed in columns 2 and 3 of
Table~\ref{tab:param}. Models whose names start with the same letter
have the same thermodynamical properties. Beam (and ambient) specific
internal energies grow from models A to D. Three different values of
the beam flow Lorentz factor have been considered for models B, C and
D. The other dependent parameters mentioned in
Section~\ref{sect:equilibrium} are displayed in columns 4-10 of
Table~\ref{tab:param}. Note that given our choice of $\rho_{0j}$, the
ambient media associated to hotter models are also hotter. The next
three columns show the longitudinal wavenumber together with
oscillation frequency and the growth rate of the most unstable
reflection mode. The following three columns display the same
quantities in the jet reference frame. Next two columns show the
perpendicular wavenumbers of linear sound waves in jet and ambient
medium respectively. The last column shows the linear growth rate of
KH instability in the jet reference frame expressed in dynamical time
units, i.e. in which time is scaled to $R_j/c_{sj}$.  All other
quantities in the table, are expressed in units of the ambient
density, $\rho_{0a}$, the speed of light, $c$, and the jet radius,
$R_j$.

  As it is apparent in Table~\ref{tab:param} growth rates tend to
increase with the specific internal energy of the beam and to decrease
with the beam flow Lorentz factor. Note that, in the jet reference
frame, models with the same thermodynamical properties tend to have
(within $\approx 10 $ \%) the same growth rates. Note also that in the
jet reference frame and in dynamical time units, all the models have
comparable (within a factor 1.5) linear growth rates.

%
\begin{table*}
\begin{center}
$
\begin{array}{|c|cc|ccccccc|ccc|ccc|ccc|}
\hline
 {\rm Model} &        \gamma &    \varepsilon_j &    \varepsilon_a &       c_{sj}  &        c_{sa} &            
    p &           \nu &          \eta &           M_j & k_{\parallel} &      \omega_r &      \omega_i  
       &{k'_\parallel} &   \omega'_{r} & \omega'_i   & k_{j\perp} &   
       k_{a\perp} & {\omega'}^{\rm dyn}_i\\
\hline
         {\rm A05} &            5 &          0.08 &          0.008 &            0.18 &          0.059 &          0.0027 &            0.11 &            0.11 &            5.47 &            0.30 &            0.20 &          0.026 &            1.32 &            7.20 &            0.13 &            7.08 &            0.53 &            0.73 \\
         {\rm B05} &            5 &          0.42 &          0.042 &            0.35 &          0.133 &          0.014 &            0.14 &            0.15 &            2.83 &            0.69 &            0.49 &          0.055 &            2.62 &            7.32 &            0.28 &            6.84 &            1.08 &            0.79 \\
         {\rm C05} &            5 &          6.14 &          0.614 &            0.55 &	  0.387 &          0.205 &            0.44 &            0.51 &            1.80 &            2.00 &            1.60 &          0.114 &            5.73 &            9.98 &            0.57 &            8.17 &            1.07 &            1.05 \\
         {\rm D05} &            5 &         60.0 &          6.000 &            0.57 &          0.544 &          2.000 &            0.87 &            0.90 &            1.71 &            2.63 &            2.18 &          0.132 &            7.02 &           11.56 &            0.66 &            9.18 &            0.24 &            1.15 \\
         {\rm B10} &           10 &          0.42 &          0.042 &            0.35 &          0.133 &          0.014 &            0.14 &            0.15 &            2.88 &            0.50 &            0.41 &          0.031 &            3.59 &           10.28 &            0.31 &            9.64 &            0.94 &            0.90 \\
         {\rm C10} &           10 &          6.14 &          0.614 &            0.55 &          0.387 &          0.205 &            0.44 &            0.51 &            1.83 &            1.91 &            1.72 &          0.055 &            9.77 &           17.67 &            0.55 &           14.72 &            1.49 &            1.01 \\
         {\rm D10} &           10 &         60.0 &          6.000 &            0.57 &          0.544 &          2.000 &            0.87 &            0.90 &            1.73 &            2.00 &            1.81 &          0.063 &            9.67 &           16.58 &            0.63 &           13.47 &            0.20 &            1.10 \\
         {\rm B20} &           20 &          0.42 &          0.042 &            0.35 &          0.133 &          0.014 &            0.14 &            0.15 &            2.89 &            0.46 &            0.39 &          0.014 &            6.51 &           18.76 &            0.28 &           17.60 &            0.90 &            0.81 \\
         {\rm C20} &           20 &          6.14 &          0.614 &            0.55 &        0.387 &          0.205 &            0.44 &            0.51 &            1.83 &            1.44 &            1.37 &          0.027 &           13.89 &           25.38 &            0.54 &           21.24 &            1.28 &            0.99 \\
         {\rm D20} &           20 &         60.0 &          6.000 &            0.57 &          0.544 &          2.000 &            0.87 &            0.90 &            1.74 &            2.00 &            1.91 &          0.029 &           18.11 &           31.43 &            0.58 &           25.68 &            0.31 &            1.01 \\

\hline
\end{array}
$
\end{center}


\caption{Equilibrium parameters of different simulated jet models
along with solutions of the dispersion relation (\ref{disprel}),
corresponding to fastest growing first reflection mode, taken as input
parameters for numerical simulations.  The meaning of the symbols is
as follows: $\gamma$: jet flow Lorentz factor; $\varepsilon$: specific
internal energy; $c_s$: sound speed; $p$p: pressure; $\nu$:
jet-to-ambient relativistic rest mass density contrast; $\eta$:
jet-to-ambient enthalpy contrast; $M_j$: internal jet Mach number;
$k_{\parallel, \perp}$: longitudinal/transverse wavenumbers;
$\omega_{r,i}$: real/imaginary part of the wave frequency.  Labels $a$
and $j$ refer to ambient medium and jet, respectively.  The primes are
used to assign wavenumber and complex frequency in the reference frame
comoving with jet.  The last column shows the linear growth rate of KH
instability in the jet reference frame expressed in dynamical time
units, i.e. in which time is scaled to $R_j/c_{sj}$. All the
quantities in the table, except the last column, are expressed in
units of the ambient density, $\rho_{0a}$, the speed of light, $c$,
and the jet radius, $R_j$.}

\label{tab:param}   
\end{table*}

  The initial numerical setup consists of a steady two-dimensional
slab jet model (see Fig.~\ref{f:sketch}). As stated above, a thin
shear layer between the ambient medium and the jet is used instead of
the vortex sheet. Due to symmetry properties, only half of the jet
($x>0$) has to be computed. Reflecting boundary conditions are imposed
on the symmetry plane of the flow, whereas periodical conditions are
settled on both upstream and downstream boundaries. The numerical grid
covers a physical domain of one wavelength along the jet (3 to 20
$R_j$; see Table~\ref{tab:param}) and 100 $R_j$ across (200 $R_j$ in
the case of models D). The size of the transversal grid is chosen to
prevent loses of mass, momentum and energy through the boundaries.
400 numerical zones per beam radii are used in the transverse
direction across the first 3 $R_j$. From this point up to the end of
the grid, 100 (200, in case of models D) extra numerical zones growing
geometrically have been added. The width growth factor between
contiguous zones is approximately 1.08 for models A, B and C and 1.04,
for models D. Along the jet, a resolution of 16 zones per beam radius
has been used. The agreement of the linear stability analysis with the
results of numerical simulations depends on the grid resolution (and
the width of the initial shear layer). The applied resolution of $400
\times 16$ grid zones per beam radius is chosen on the base of several
tests, which are presented in detail in Appendix A.

  The steady model is then perturbed according to the selected mode,
with an absolute value of the pressure perturbation amplitude inside
the beam of $p_j^{\pm}=10^{-5}$. This means that those models with the
smallest pressure, like model A, have relative perturbations in
pressure three orders of magnitude larger than those with the highest
pressures, D. However this difference seems not to affect the linear
and postlinear evolution.

\section{Results \label{sect:results}}

  Following the behaviour of simulated models we find that the evolution
of the perturbations can be divided into a linear phase, a saturation
phase and mixing phase. This section is devoted to describing the
properties of the linear and saturation phases in our models focusing
on the influence of the basic parameters. The remaining fully
nonlinear evolution will be discussed in Paper II. Our description
shares many points with that of Bodo et al. (1994) for the case of
classical jets.

  In order to illustrate the growth of perturbations and determine the
duration of the linear and saturation phases in our simulations, we
plot in Fig.~\ref{fig:amplitudes} the amplitudes of the perturbations
of the longitudinal and transversal velocities, inside the jet and in
the jet reference frame, together with the pressure oscillation
amplitude. We plot also the growth of the imposed eigenmodes resulting
from the linear stability analysis. Both the velocity perturbations
are transformed from the ambient rest frame to the unperturbed jet
rest frame using the Lorentz transformation rules for velocity
components, given by Eqn.~(\ref{eq:lorentz}).

  We define the characteristic times in the following way. During the
linear phase the ratios of the exponentially growing amplitudes of
pressure, longitudinal velocity and transversal velocity remain
constant by definition.  We define the end of the linear phase
($t_{\rm lin}$) as the moment at which one of quantities starts to
depart from the initial exponential growth. Within the set of our
models the first quantity to break the linear behaviour is the
longitudinal velocity perturbation.

  Later on the transversal velocity saturates, i.e. stops growing at the
saturation time ($t_{\rm sat}$).  We call the period between $t_{\rm
lin}$ and $t_{\rm sat}$ the saturation phase. We find also that the
pressure perturbation amplitude reaches a maximum. This moment is
denoted by $t_{\rm peak}$. In Paper II we will see that this peak
announces the entering of the fully nonlinear regime. The choice of
$t_{\rm lin}$, $t_{\rm sat}$, $t_{\rm peak}$ has been illustrated in
Fig.~\ref{fig:amplitudes} (top panel).

  Table~\ref{tab:phases} collects times of the linear and saturation
phases in the different models (columns 2-4). The last column shows the
saturation time in dynamical units and in the jet reference frame. The
change of reference frame eliminates the effects coming from the jet
flow Lorentz factor that stretches out the rhythm of evolution in the
ambient rest frame. Dynamical time units are adapted to the
characteristic time of evolution of each model. Hence this change of
units and reference frame allows us to compare the relevant scales of
evolution of all the models directly.

%
\begin{table}

\begin{center}
$
\begin{array}{|c|ccc|c|}
\hline
{\rm Model} & t_{\rm lin} & t_{\rm sat} & t_{\rm peak} 
            &  {t'}_{\rm sat}^{\rm dyn}\\
\hline
{\rm A05}   & 180   & 380   & 380   & 13.69   \\
{\rm B05}   & 125   & 200   & 205   & 13.99   \\
{\rm C05}   & 100   & 125   & 130   & 13.74   \\
{\rm D05}   & 105   & 120   & 130   & 13.71   \\
\hline
{\rm B10}   & 235   & 380   & 385   & 13.29   \\
{\rm C10}   & 210   & 245   & 250   & 13.46   \\
{\rm D10}   & 180   & 225   & 225   & 12.86   \\
\hline
{\rm B20}   & 450   & 760   & 780   & 13.29   \\
{\rm C20}   & 270   & 645   & 775   & 17.72  \\
{\rm D20}   & 350   & 480   & 500   & 13.71   \\
\hline
\end{array}
$
\end{center}

\caption{Times for the different phases in the evolution of the
perturbed jet models. In columns 2-4 we show: $t_{\rm lin}$ - end of
linear phase (the ratios of amplitudes of the different quantities are
not constant any longer), $t_{\rm sat}$ - end of saturation phase (the
amplitude of the transverse speed perturbation reaches its maximum),
$t_{\rm peak}$ - the peak in the amplitude of the pressure
perturbation is reached. (see Fig. \ref{fig:amplitudes}). Note that,
as a general trend, $t_{\rm lin} < t_{\rm sat} \leq t_{\rm peak}$ .
In columns 5 we show the saturation time in dynamical time units and
in the jet reference frame.}
\label{tab:phases}   

\end{table}
%

\subsection{Linear phase}
\label{ss:linear}

  In the linear phase the ratios of oscillation
amplitudes of different dynamical variables (density, pressure and
velocity components) are constant. We have used this property to
identify the end of the linear phase with the time $t_{\rm lin}$ at
which one of the variables deviates from linear growth in a
systematic way.  Note that our definition is different from that of
Bodo et al. (1994) who associate the end of the linear phase with the
formation of the first shocks inside the jet.

  We note that in our simulations, during the linear phase, the growth
of perturbations of each dynamical variable follows the predictions of
the linear stability analysis with relatively good accuracy. On
average the growth rates measured for all our numerical experiments
are about 20 \% smaller than corresponding growth rate resulting from
the linear stability analysis. This small discrepancy may be
partially a result of the application of the shearing layer in
simulations and the vortex sheet approximation in the linear stability
analysis, and also to a lack of transversal resolution in the numerical
simulations (see Appendix A).

  In all cases the amplitude of the longitudinal velocity oscillation
is the first quantity to stop growing. The reason for the limitation
of the velocity oscillations is obvious: the oscillations of velocity
components (corresponding to sound waves propagating in the jet
interior) cannot exceed the speed of light. This limitation (specific
to relativistic dynamics) is easily noticeable in the jet reference
frame, but it is obscured by the Lorentz factor (in
the first and second power) in the reference frame of the ambient
medium, as it follows from the formulae (\ref{lorentz-pert}).

\begin{figure*}
\hspace{0.4cm} \psfig{file=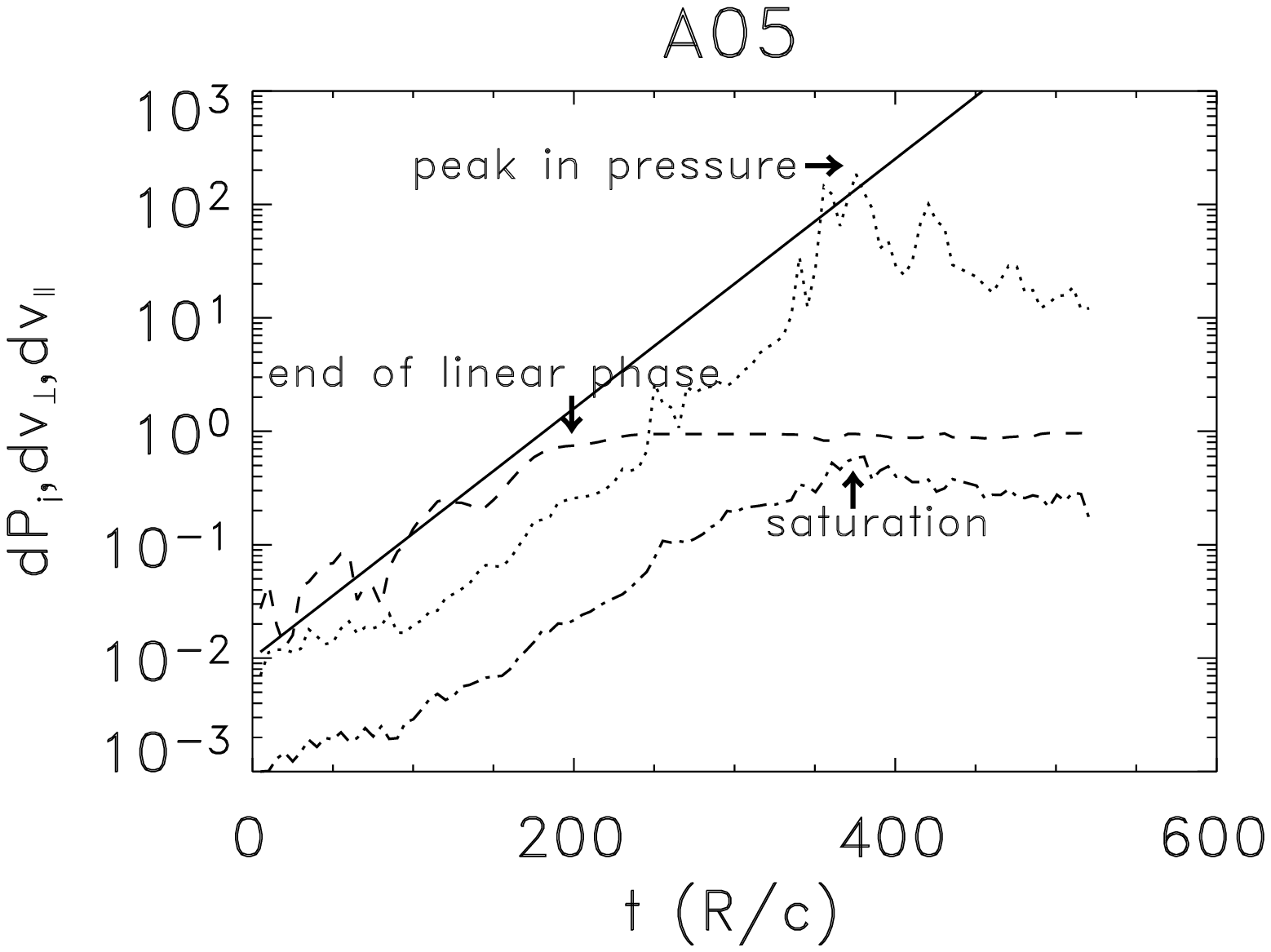,width=0.3 \textwidth,angle=0,clip=} \\
\centerline{
\psfig{file=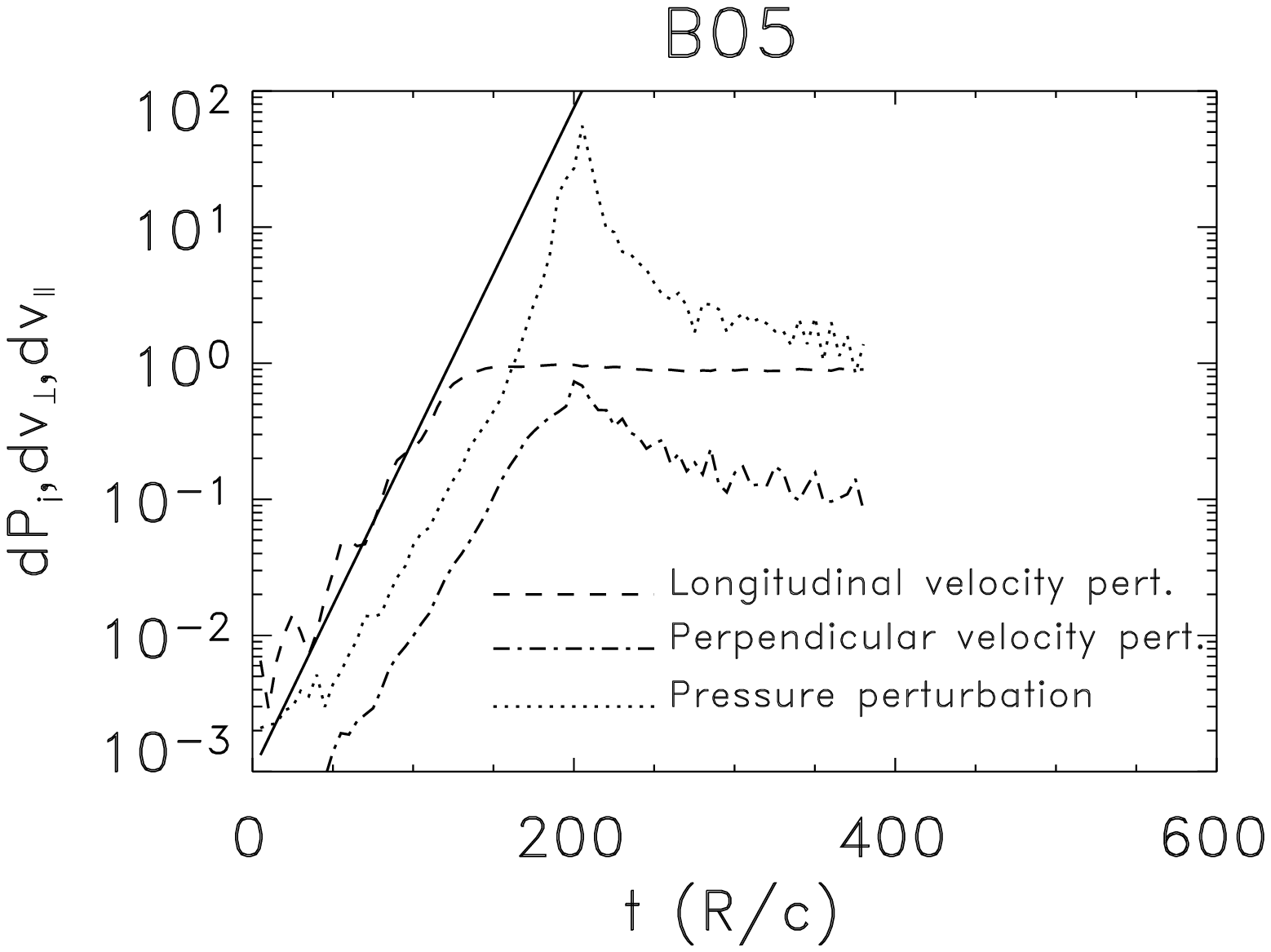,width=0.3 \textwidth,angle=0,clip=} \quad 
\psfig{file=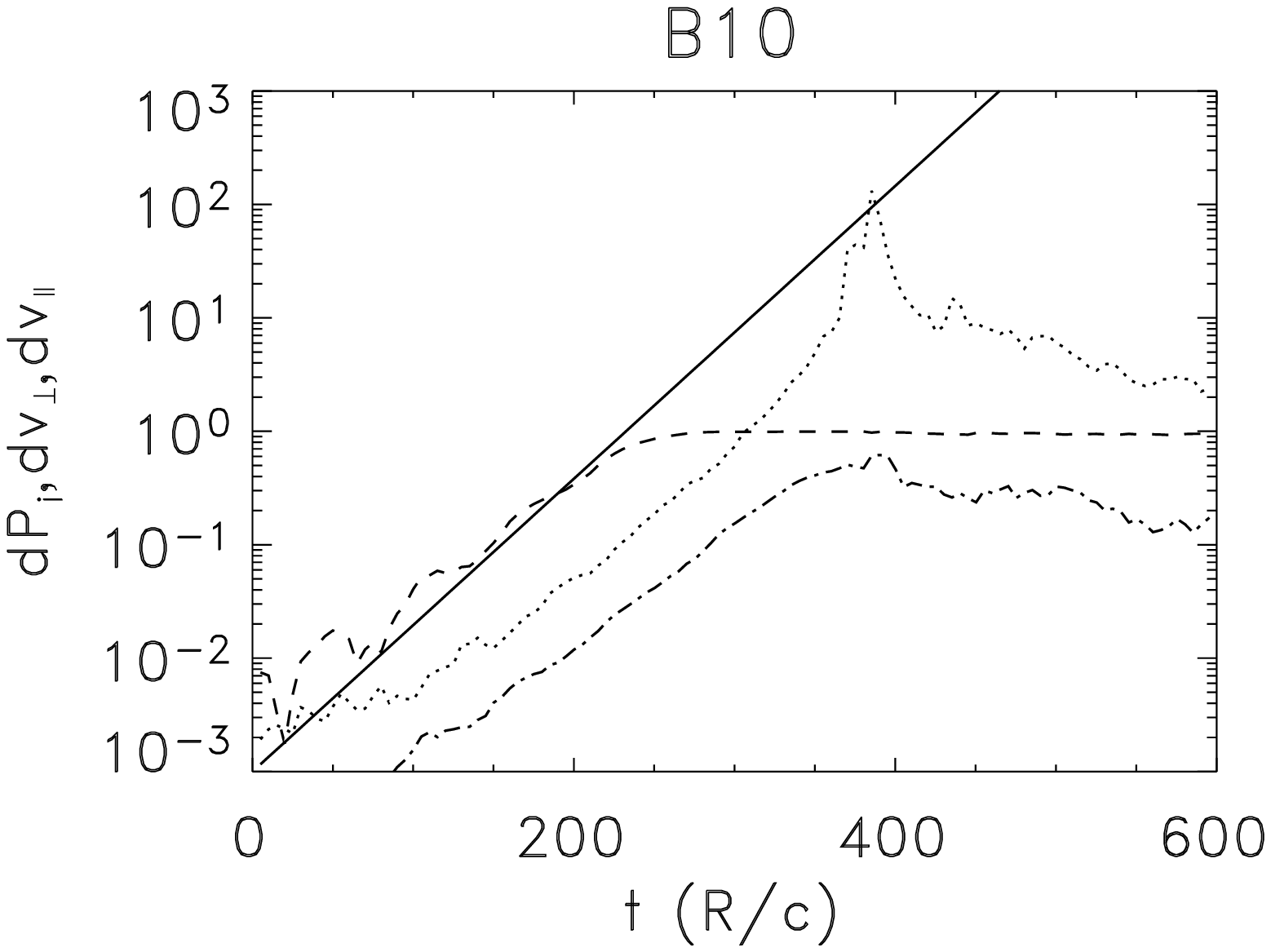,width=0.3 \textwidth,angle=0,clip=} \quad 
\psfig{file=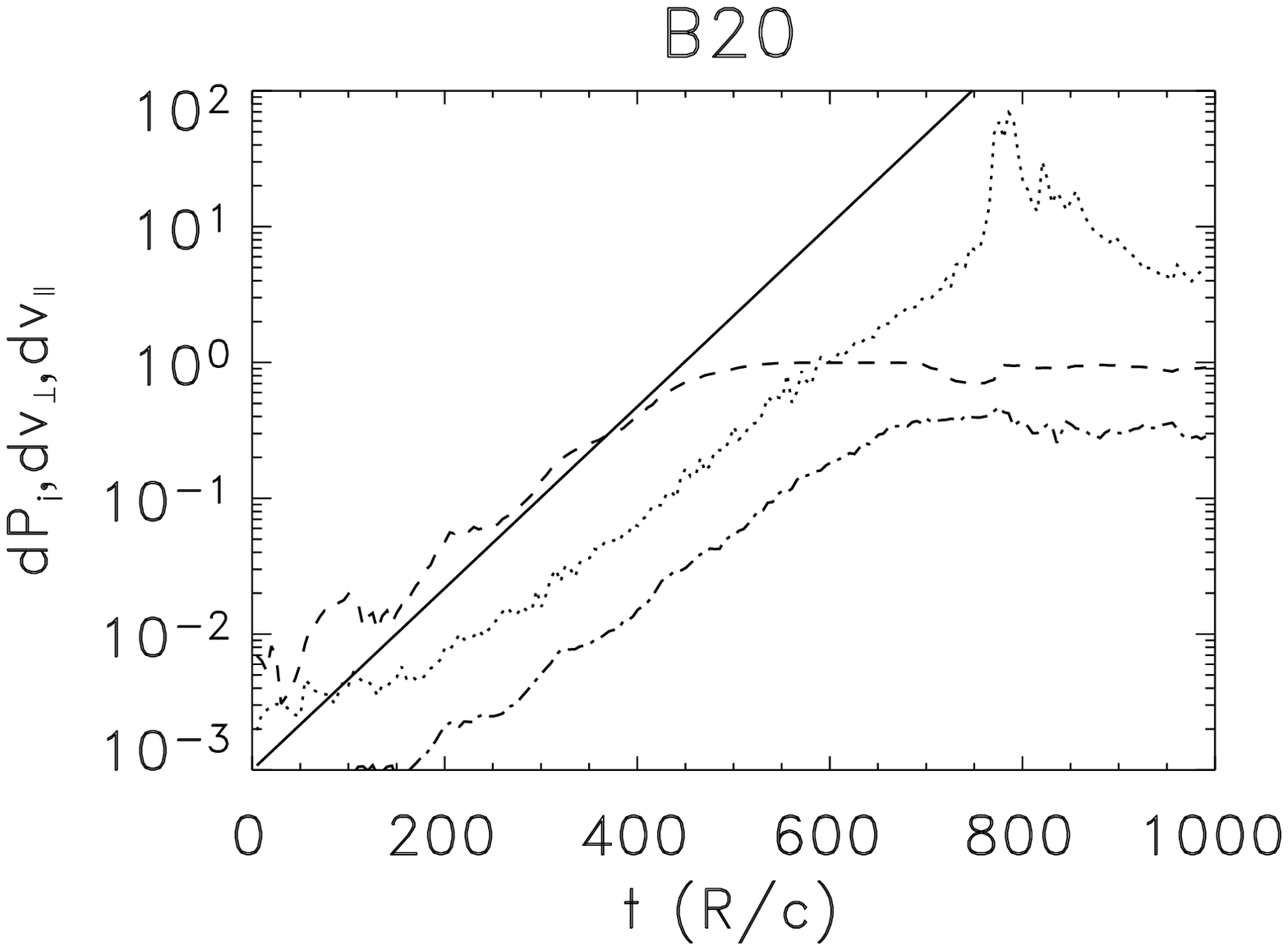,width=0.3 \textwidth,angle=0,clip=} 
}
\centerline{
\psfig{file=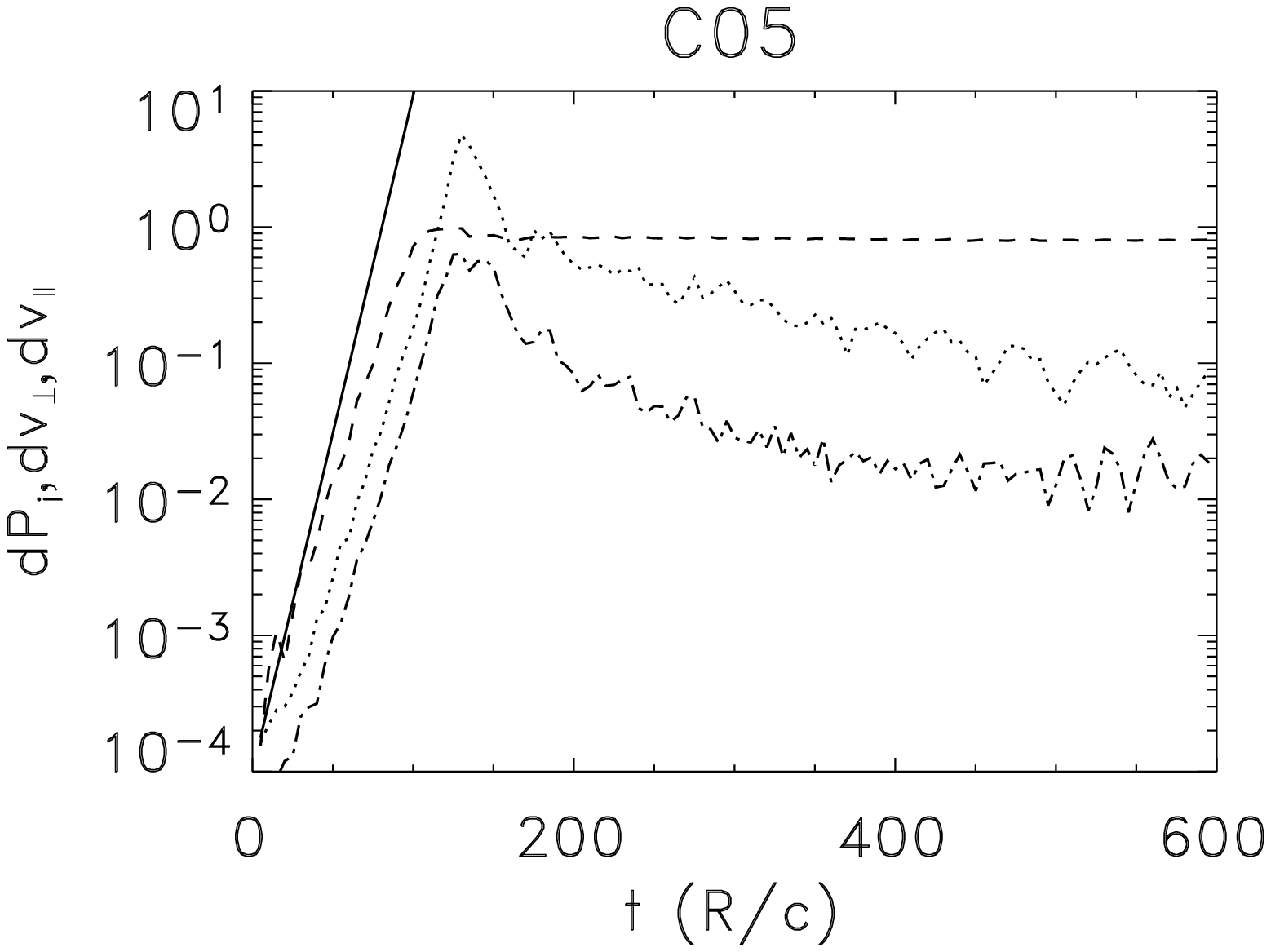,width=0.3 \textwidth,angle=0,clip=} \quad 
\psfig{file=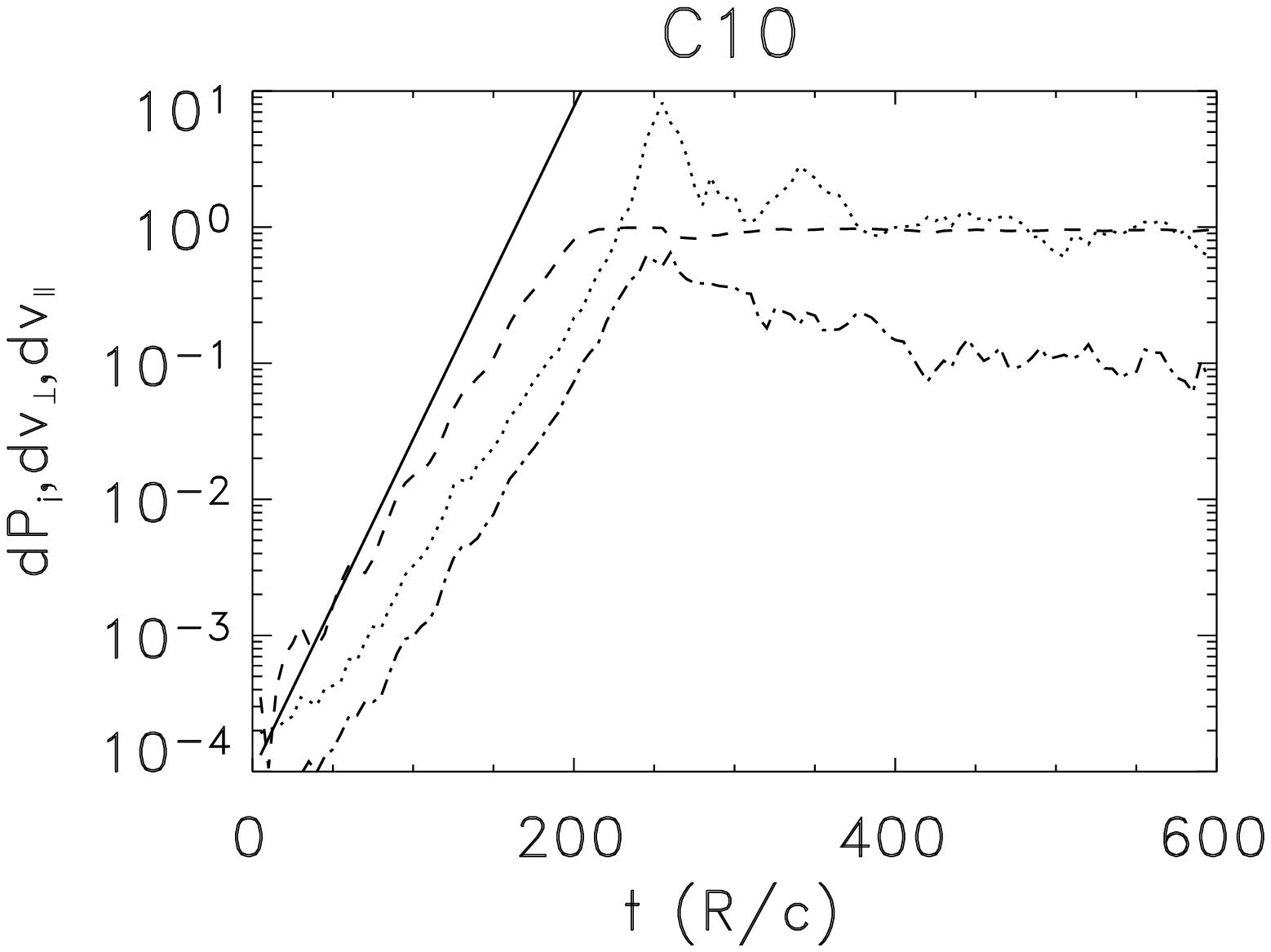,width=0.3 \textwidth,angle=0,clip=} \quad 
\psfig{file=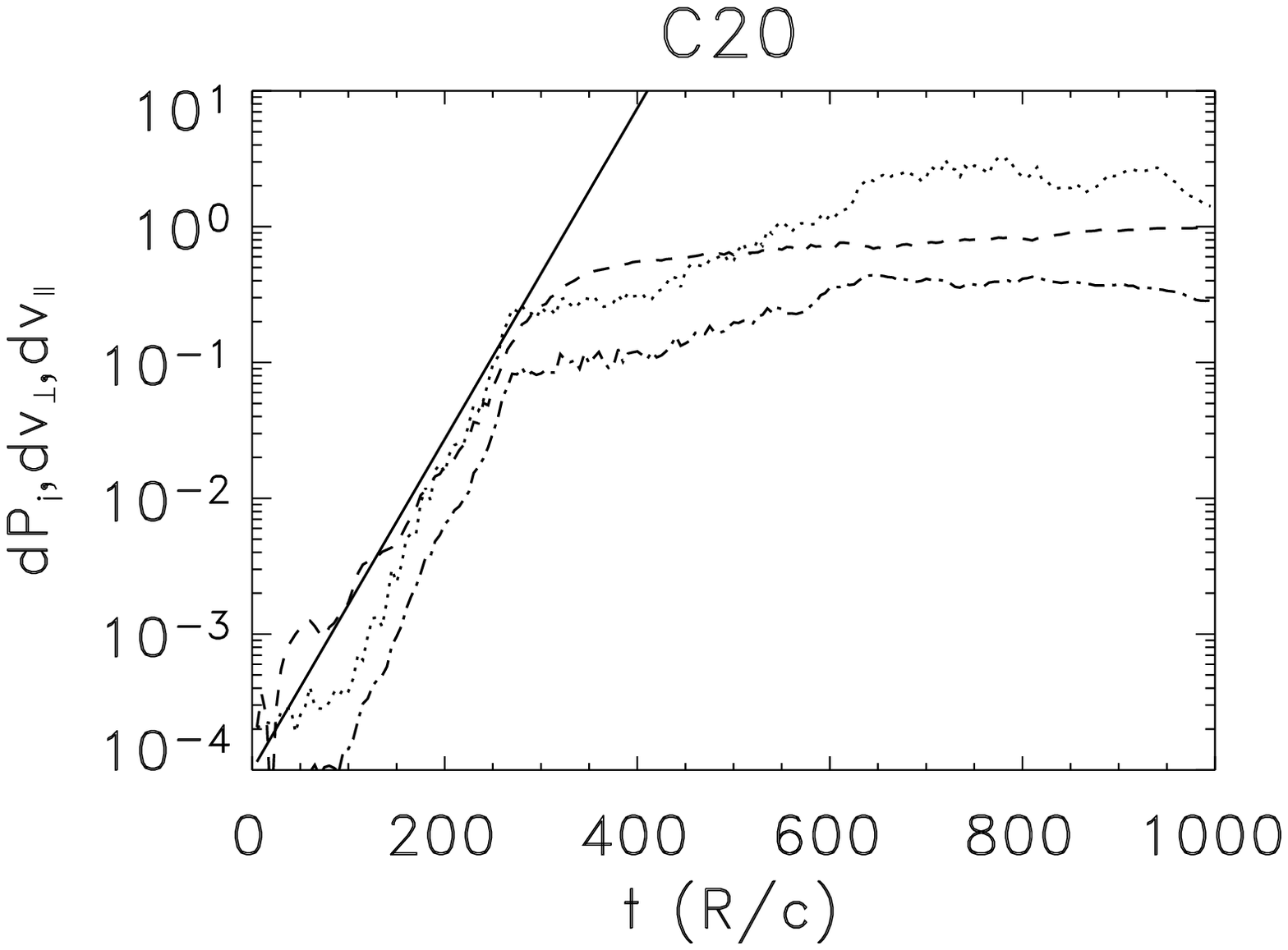,width=0.3 \textwidth,angle=0,clip=} 
}
\centerline{
\psfig{file=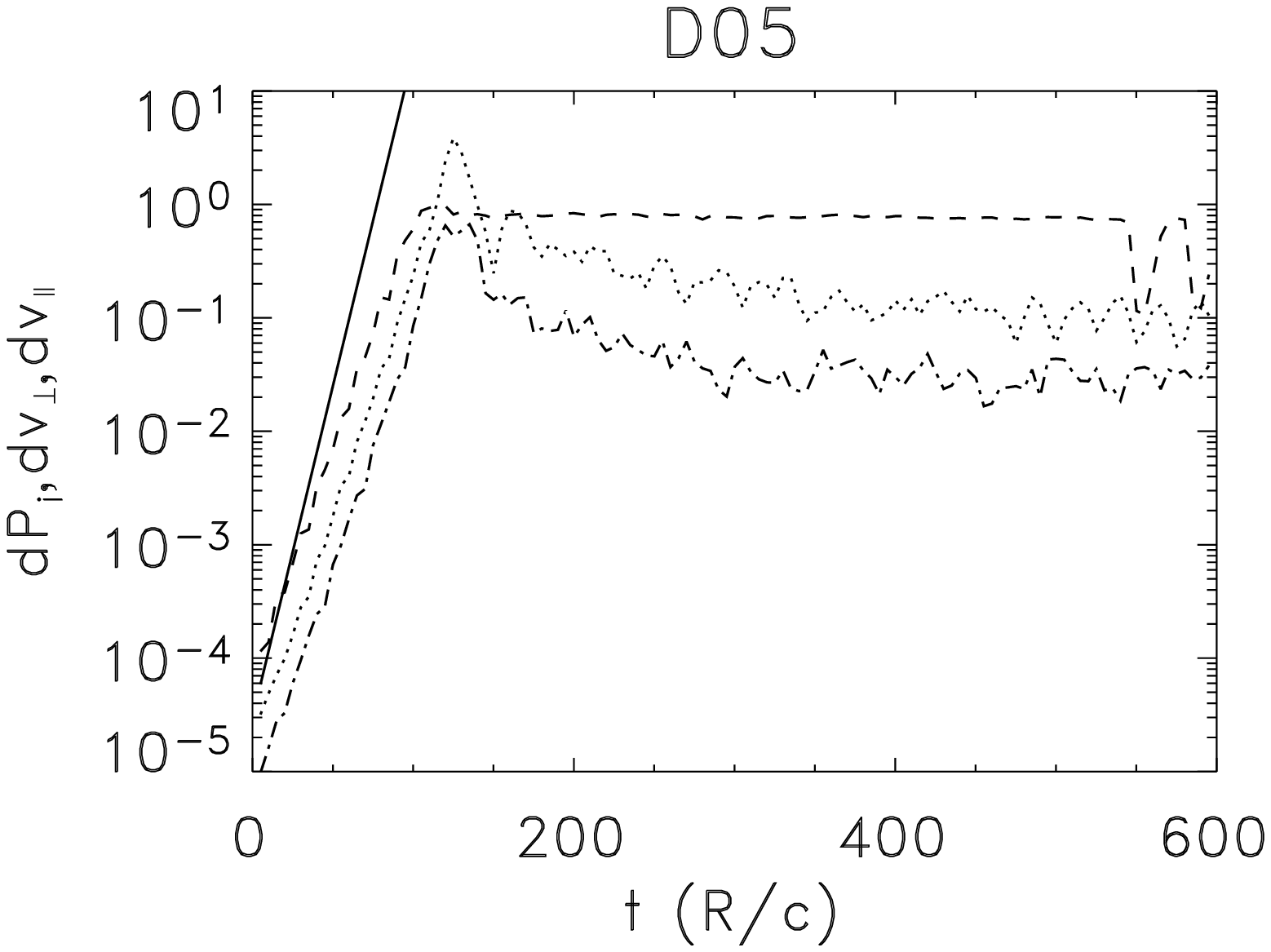,width=0.3 \textwidth,angle=0,clip=} \quad 
\psfig{file=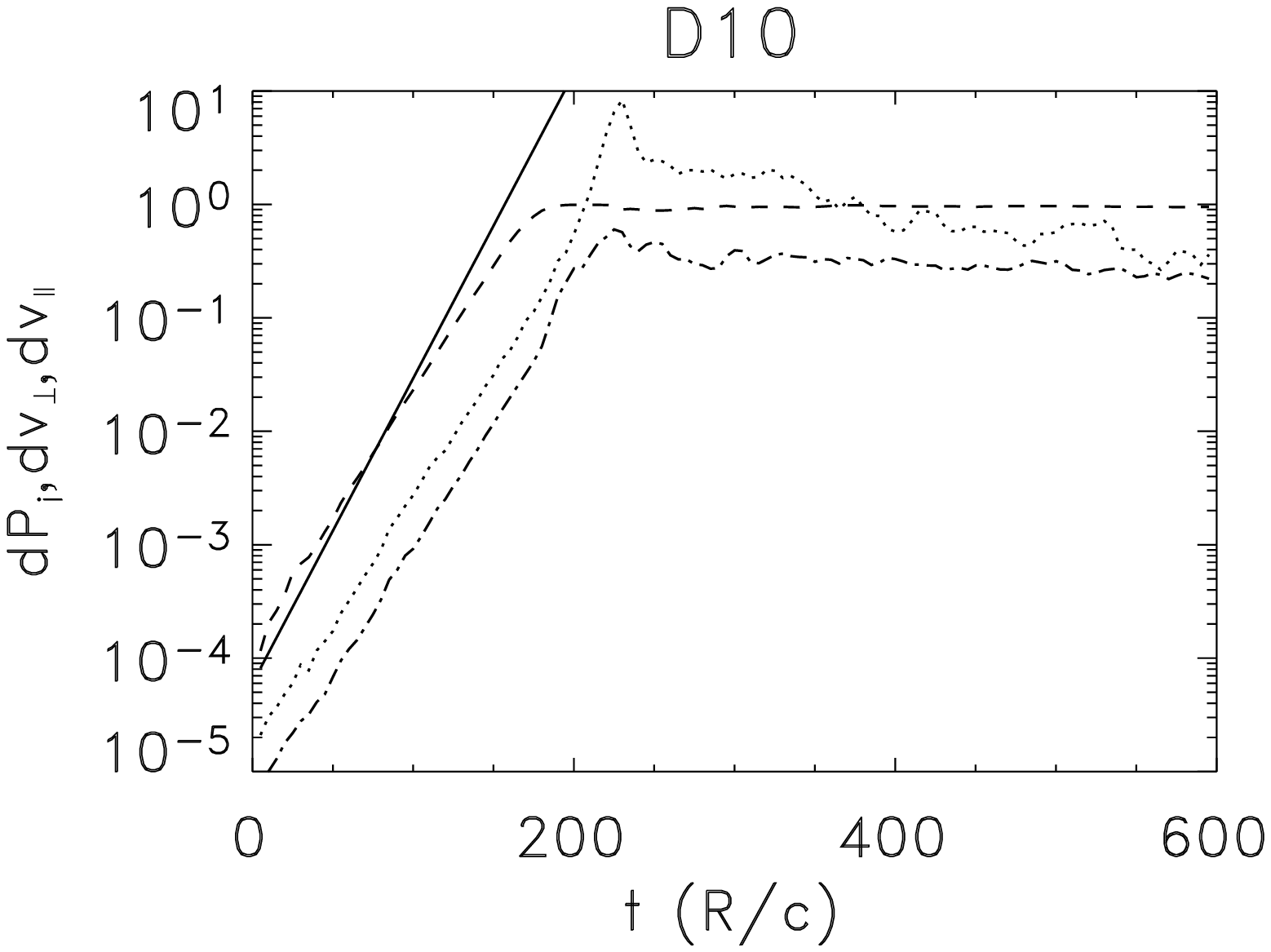,width=0.3 \textwidth,angle=0,clip=} \quad 
\psfig{file=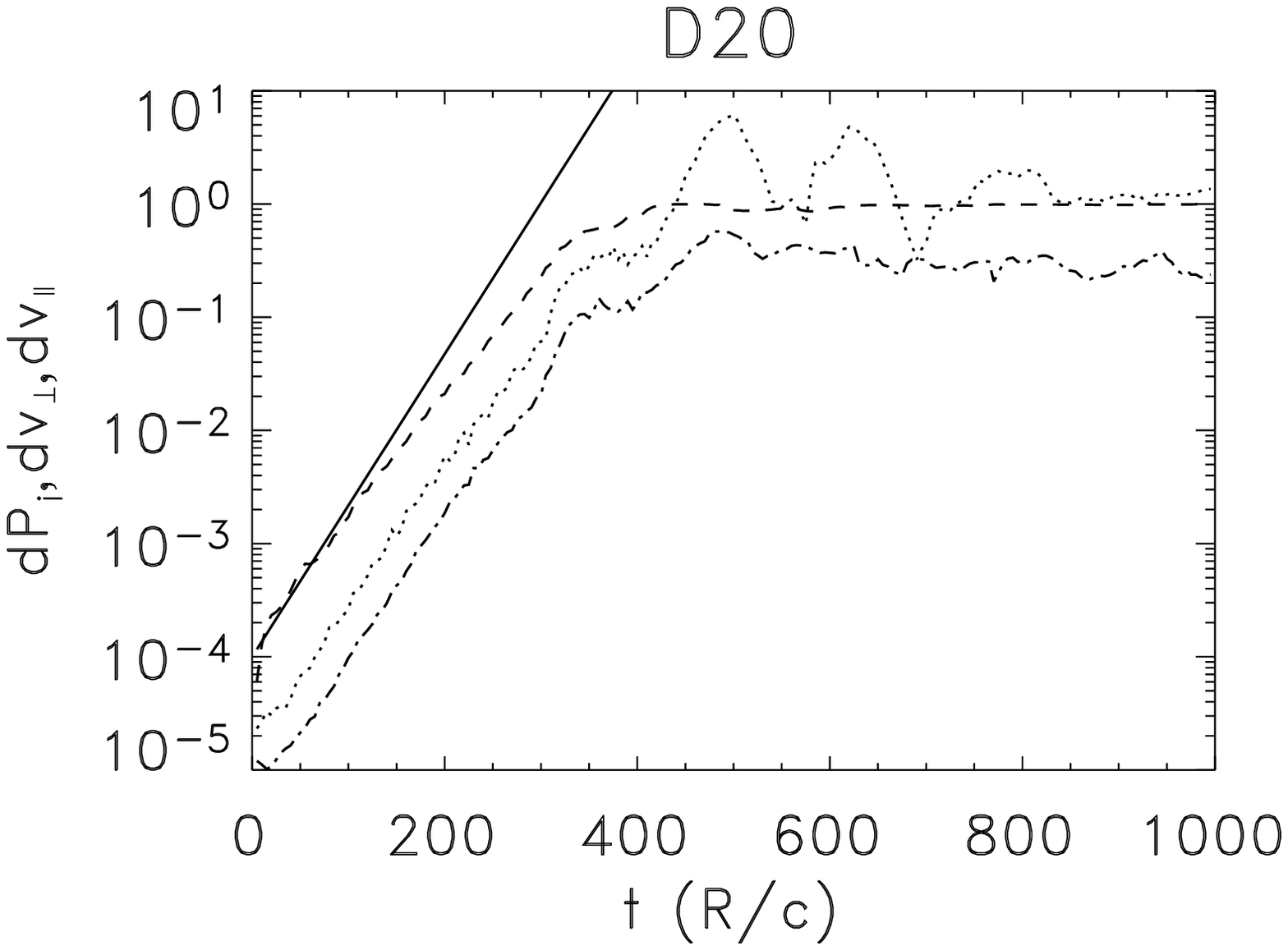,width=0.3 \textwidth,angle=0,clip=} 
}

\caption{Evolution of the relative amplitudes of perturbations. Dotted
  line: pressure perturbation ($(p_{max}-p_0)/p_0$). Dashed line:
  longitudinal velocity perturbation in the jet reference frame
  ($0.5\,(v'_{z,max}-v'_{z,min})$). Dash-dotted line: perpendicular
  velocity perturbation in the jet reference frame
  ($0.5\,(v'_{x,max}-v'_{x,min})$). The search for maximum ($(p_{max}$,
  $(v'_{x,max}$, $(v'_{z,max}$) and minimum ($v'_{x,min}$, $(v'_{z,min}$)
  values have been restricted to those numerical zones with jet mass
  fraction larger than 0.5. Solid line: linear analysis prediction
  for the growth of perturbation. Note that abscissae in the last
  column plots extend up to $t= 1000 R_j/c$ and that ordinate values
  adapt to fit the scale of each plot. Arrows in the plot of
  Model A05 point to specific stages of evolution used to define
  $t_{\rm lin}$, $t_{\rm sat}$ and $t_{\rm peak}$ (see text).}
\label{fig:amplitudes}
\end{figure*}      
%

  As seen in Fig.~\ref{fig:amplitudes}, model C20 evolves distinctly
from the other models: at $t = 270$, the linear phase ends before the
saturation of longitudinal speed. See the middle panels of
Fig.~\ref{fig:fcd20s} (especially the pressure panel) which could be
responsible for the peculiar evolution of model C20. Note also
that model C20 is the only one that has a theoretical perturbation
growth rate smaller than the numerical one (probably associated with
the excited short wavelength mode).

\begin{figure*}
\centerline{
\psfig{file=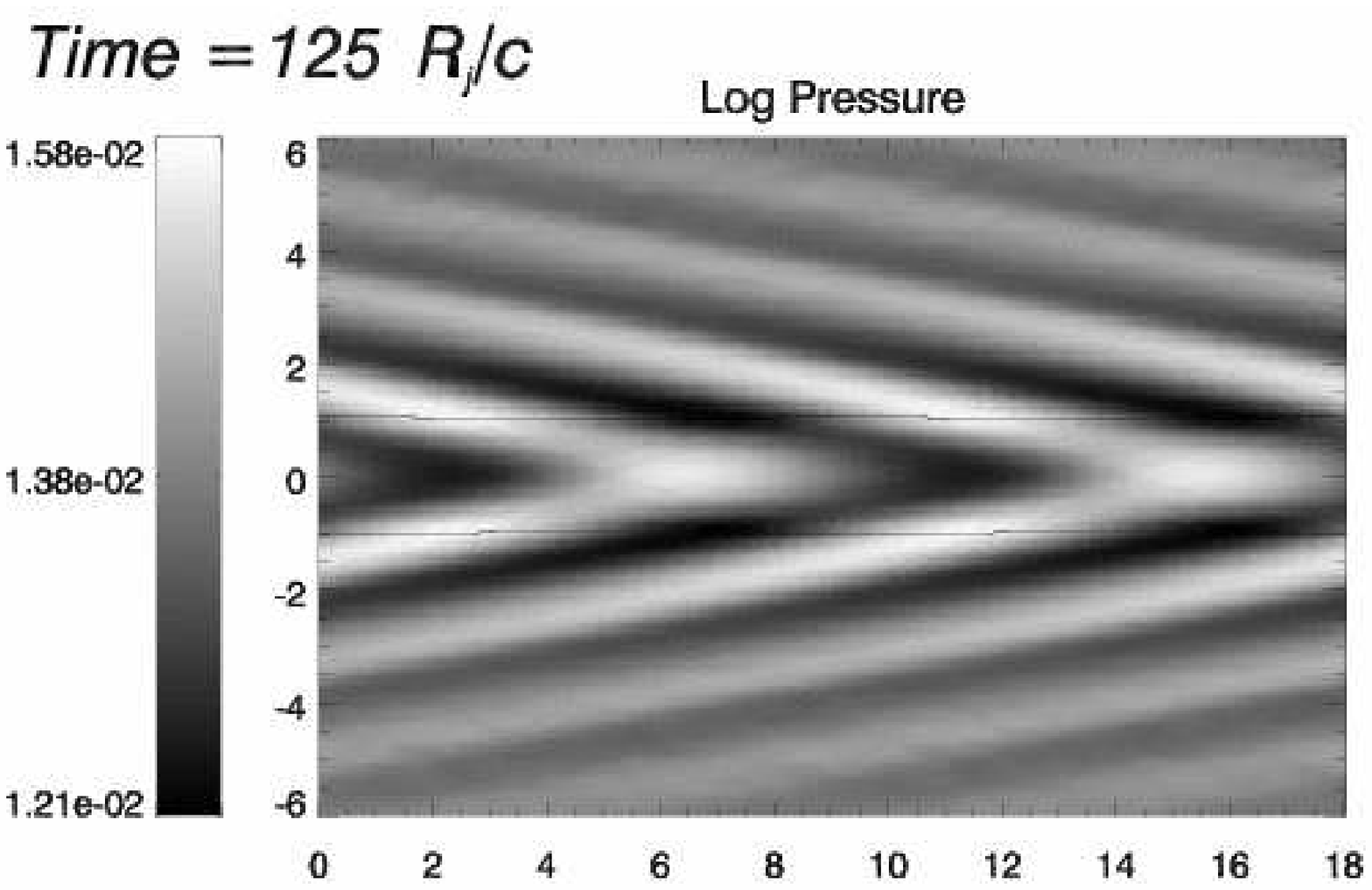,width=0.45 \textwidth,angle=0,clip=0} \quad
\psfig{file=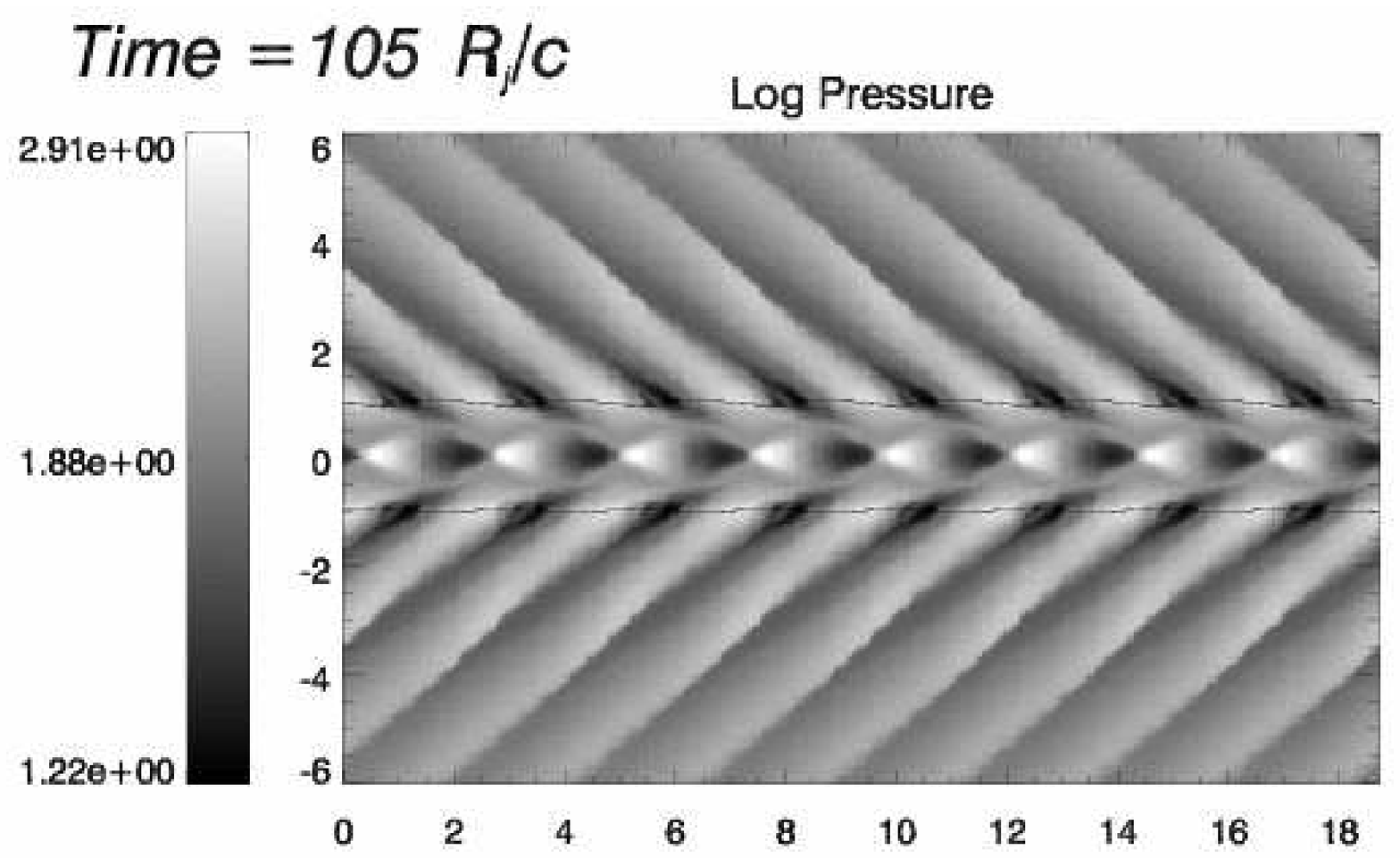,width=0.45 \textwidth,angle=0,clip=0}
}

\caption{Pressure distribution at the end of linear phase for models
  B05 (left panel) and D05 (right panel). The continuous line
  corresponds to jet mass fraction equal to 0.5 and serves to
  distinguish jet and ambient media.}
\label{fig:prepert}
\end{figure*}

  At the end of the linear phase, the values of quantities like
density, pressure and flow Lorentz factor are still very close to the
corresponding background values (the perturbation in pressure is
between 10 \% and 50\% of the background pressure in all the models).
Figure~\ref{fig:prepert} shows the distribution of pressure at the end
of the linear phase for two representative models B05 and D05.

%
%

\subsection{Saturation phase}

  As it has been mentioned, the end of the linear phase coincides with
the limitation of the longitudinal oscillation velocity. At time
$t_{\rm lin}$ the transversal velocity component is smaller than the
speed of light, by an order of magnitude approximately, so the
transversal velocity perturbation has still room to grow. We 
defined $t_{\rm sat}$ as the time corresponding to the
saturation of the transversal velocity and saturation phase as the
period between $t_{\rm lin}$ and $t_{\rm sat}$.

  We find that the longitudinal velocity perturbation amplitude
reaches almost the speed of light (more than $0.9 c$) while the
transversal perturbation amplitude stops its growth at the level of
approximately $0.5c$ for all presented simulations. 
This can be explained in the following way. We remind
that the eigenmodes are built of oblique sound waves overreflected at
the jet/ambient medium interface. This means that the amplitude of the
reflected wave is larger than the amplitude of the incident one. Sound
waves are longitudinal waves, so the gas oscillation velocity can be
split into the longitudinal and transversal components separately for
the incident and reflected waves.  Let us consider locally the
incident wave and the reflected wave in the jet medium, in a fixed
point close to the jet boundary. Then the longitudinal velocity
components of the incident and reflected waves sum in phase, while
the transversal ones sum in counterphase. This means that while the
amplitude of the total longitudinal velocity oscillation component
grows and approaches the speed of light, the oscillation amplitude of
the total transversal component is a difference of the two velocities, each
smaller than the speed of light. Therefore the oscillation amplitude
of the total transversal component has to take values significantly
smaller than the longitudinal one.

  The duration of the saturation phase depends on the Lorentz factor and
the specific internal energy with a tendency to increase with the
former and to decrease with the latter (exception made of models C10
or D10 and C20). It ranges between a few tens of (absolute) time units
to a few hundreds.

  We note that the saturation times expressed in dynamical time units
and in the jet reference frame (see Table~\ref{tab:phases}) are almost
equal for all models.  This similarity can be explained by the
following argument. First, the amplitude of
pressure perturbations $p_1^\pm=10^{-5}$ is the same for all
simulations. Second, the amplitudes of pressure and transversal
velocity perturbations are related by formula
(\ref{pj-vjperp}). The perturbation of perpendicular velocity is then
transformed to the jet reference frame following (\ref{lorentz-pert}).
Following the results of numerical experiments, the transversal
velocity grows with the linear growth rate until the transversal
velocity perturbation reaches the upper limit, i.e. $|v_{j\perp}^{+}|
\exp (\omega_i t_{\rm sat}) \simeq 0.5c $. If we express $t_{\rm sat}
$ in dynamical time units then in the jet reference frame

\begin{equation}
{t'}_{\rm sat}^{\rm dyn}  \simeq \frac{\gamma}{c_{sj}  \omega_i}
  \log \left(\frac{c \Gamma_j}{2p_1^+}
  \left|\frac{\omega'}{k_{j\perp}}\right|\right) \label{tsat}.
\end{equation}

\noindent 
Since the term $\omega'/k_{j\perp}$ under the logarithm is close to
the jet sound speed and is varying only slightly in between models,
while the other factors under the logarithm are constants, the
saturation time is proportional to the inverse growth rate. From the
linear stability analysis, the latter one measured in the jet
reference frame and expressed in dynamical time units is almost equal
for all the models (see last column in Table~\ref{tab:param}).
Substitution of numbers into the formula (\ref{tsat}) provides
$t'_{\rm sat}$ in the range $10 \div 15$, which is consistent with the
value 13.5 in Table~\ref{tab:phases}.  However, the remarkable
convergence of saturation times for all the models is difficult to
explain in view of the additional randomness in the evolution of
perturbations, which is apparent in Fig.~\ref{fig:amplitudes}.

\begin{figure*} 
\centerline{
\psfig{file=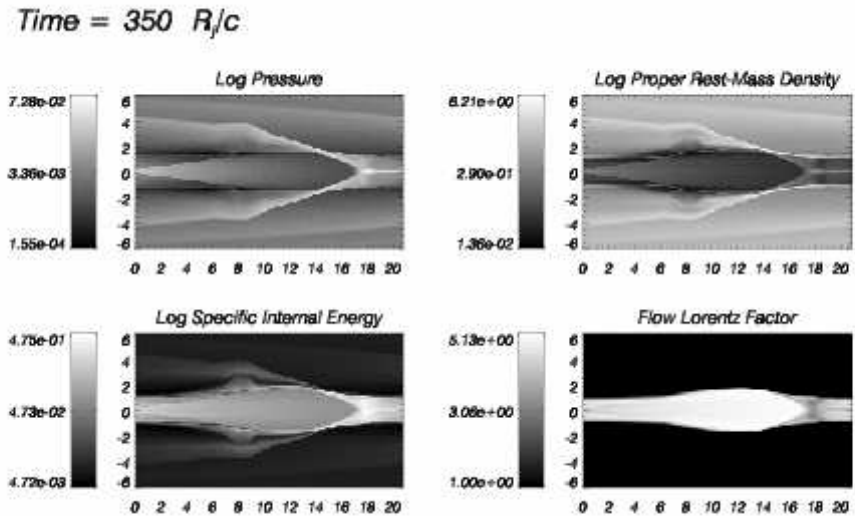,width=0.8\textwidth,angle=0,clip=0} 
}
\caption{Snapshot around saturation of logarithmic maps of pressure, 
rest-mass density and specific internal energy and non-logarithmic
Lorentz factor for model A05.} 

\label{fig:fab05s}

\end{figure*}

\begin{figure*}
\centerline{
\psfig{file=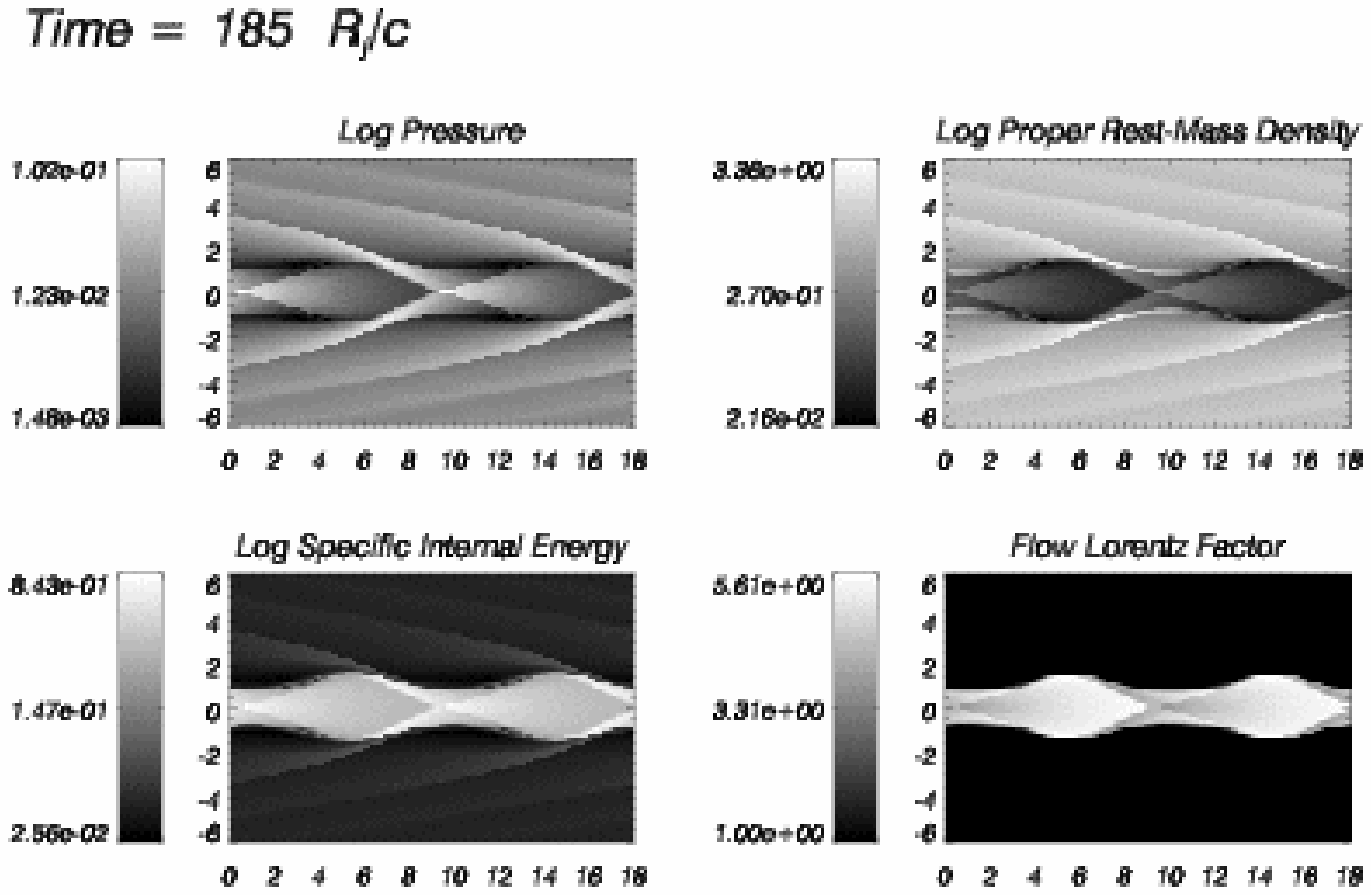,width=0.8\textwidth,angle=0,clip=0} 
}
\centerline{
\psfig{file=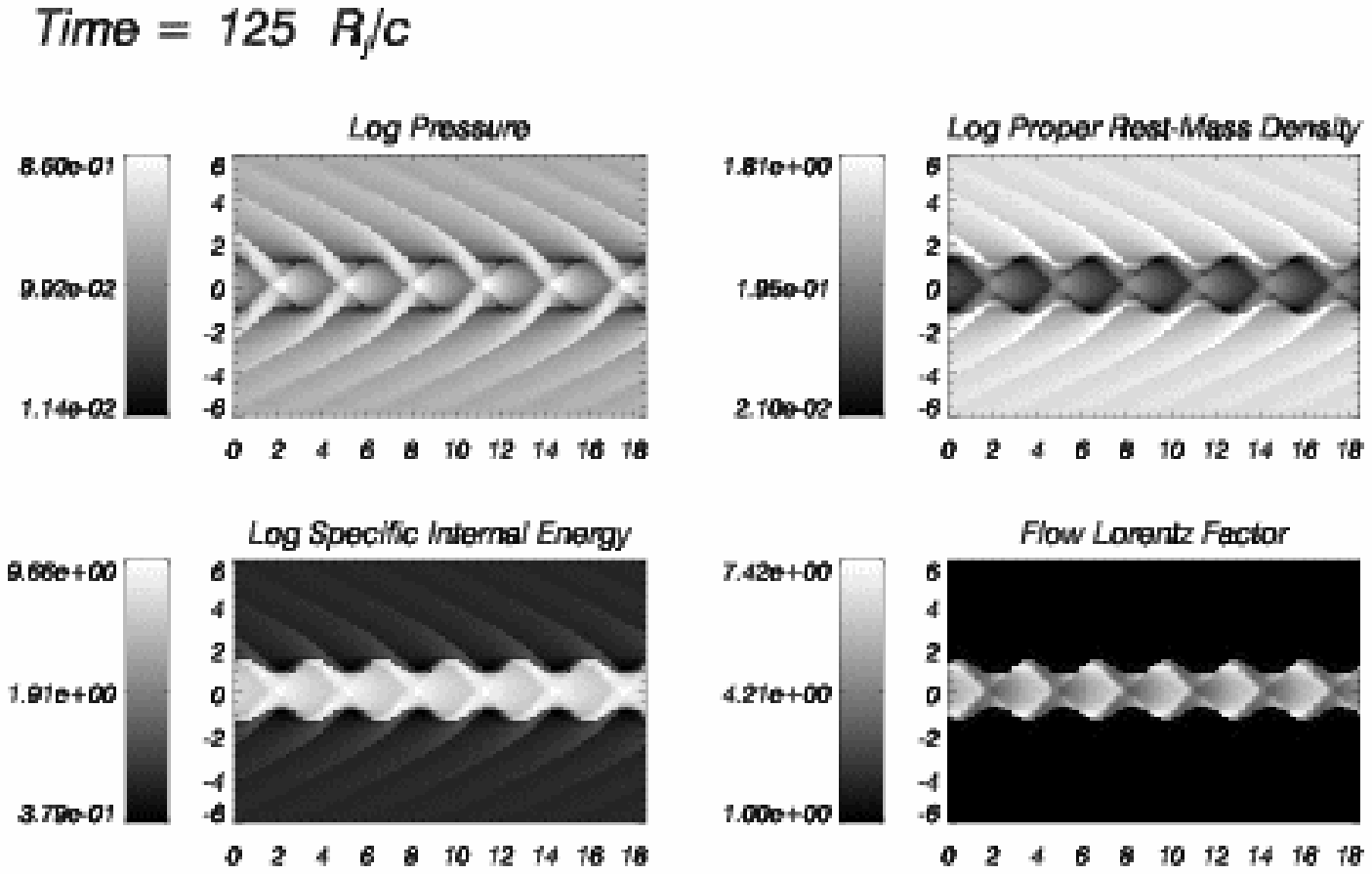,width=0.8\textwidth,angle=0,clip=0} 
}
\centerline{
\psfig{file=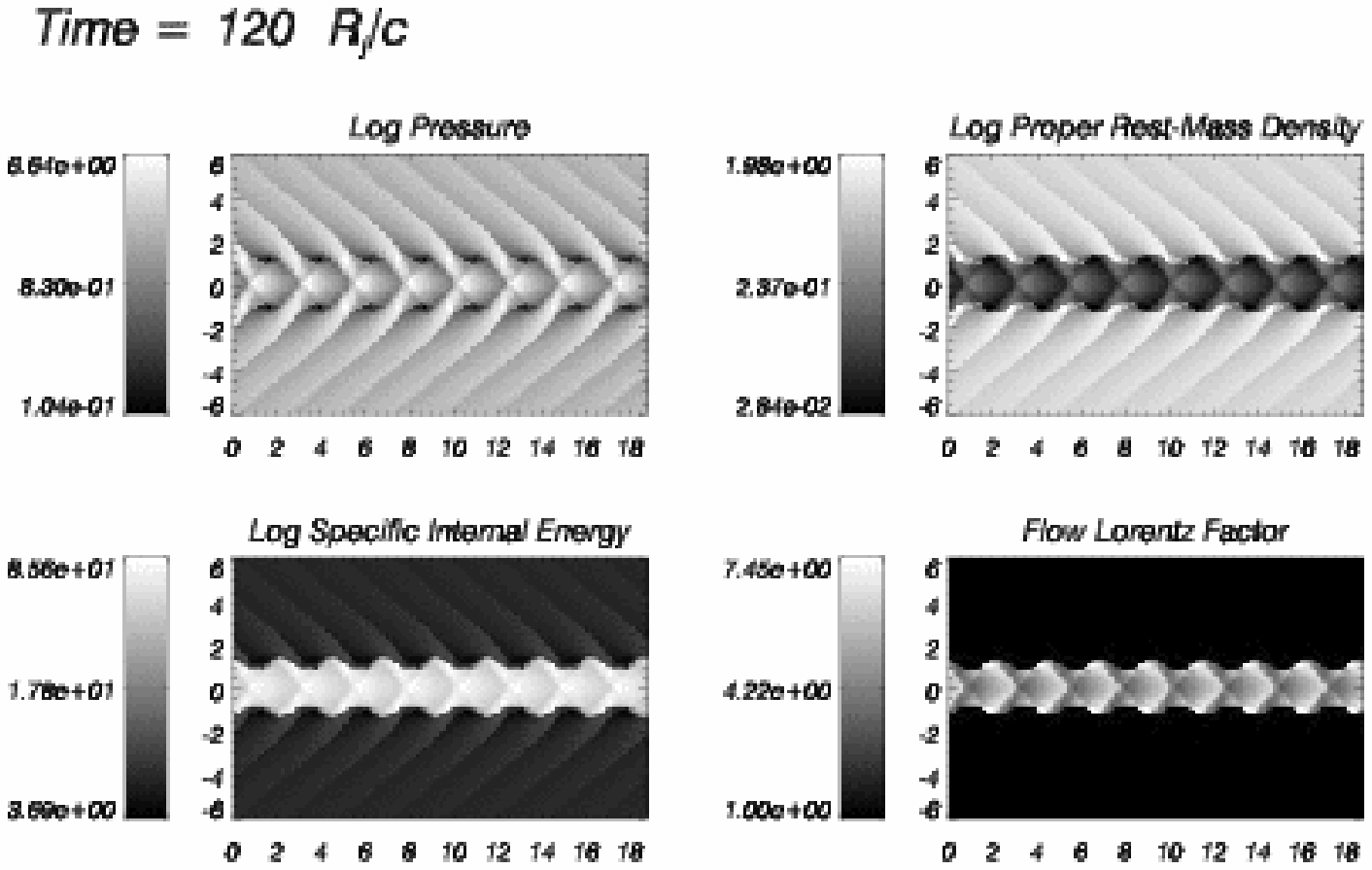,width=0.8\textwidth,angle=0,clip=0}
}
\caption{Same as Fig. (\ref{fig:fab05s}) for models B05 (upper), 
C05 (middle) and D05 (lower).} 
\label{fig:fcd05s}

\end{figure*}

\begin{figure*}
\centerline{
\psfig{file=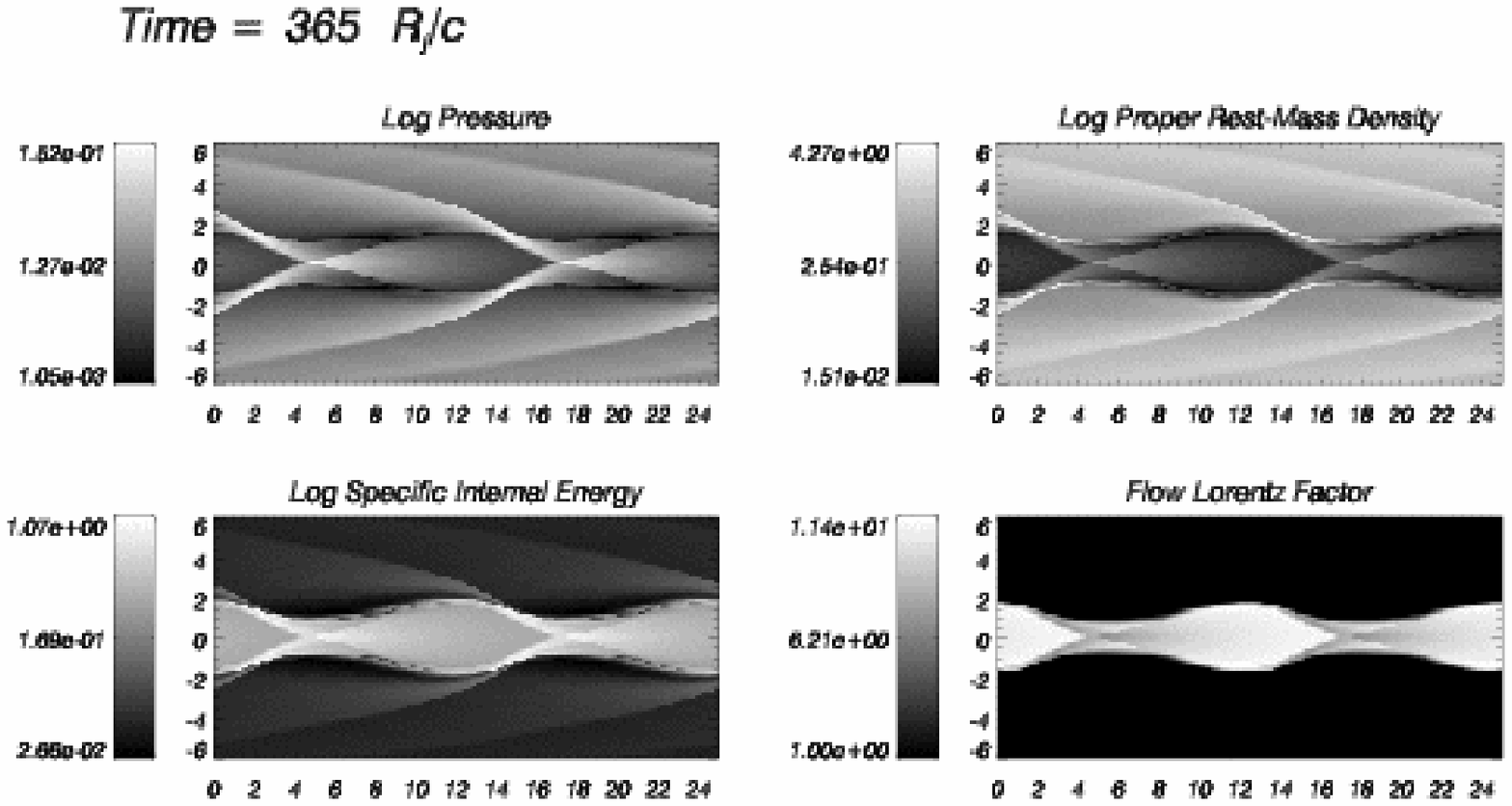,width=0.8\textwidth,angle=0,clip=0} 
}
\centerline{
\psfig{file=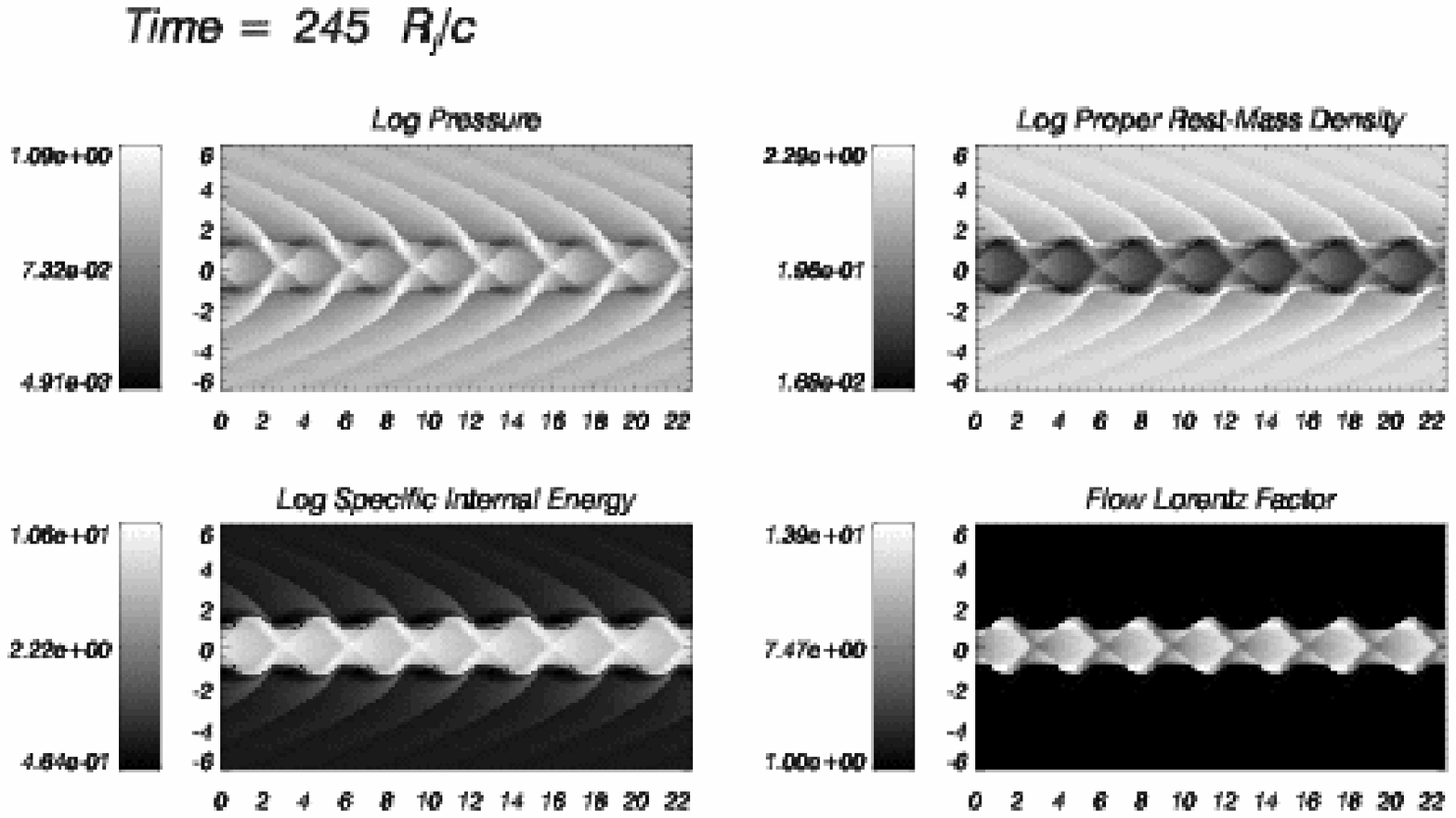,width=0.8\textwidth,angle=0,clip=0} 
}
\centerline{
\psfig{file=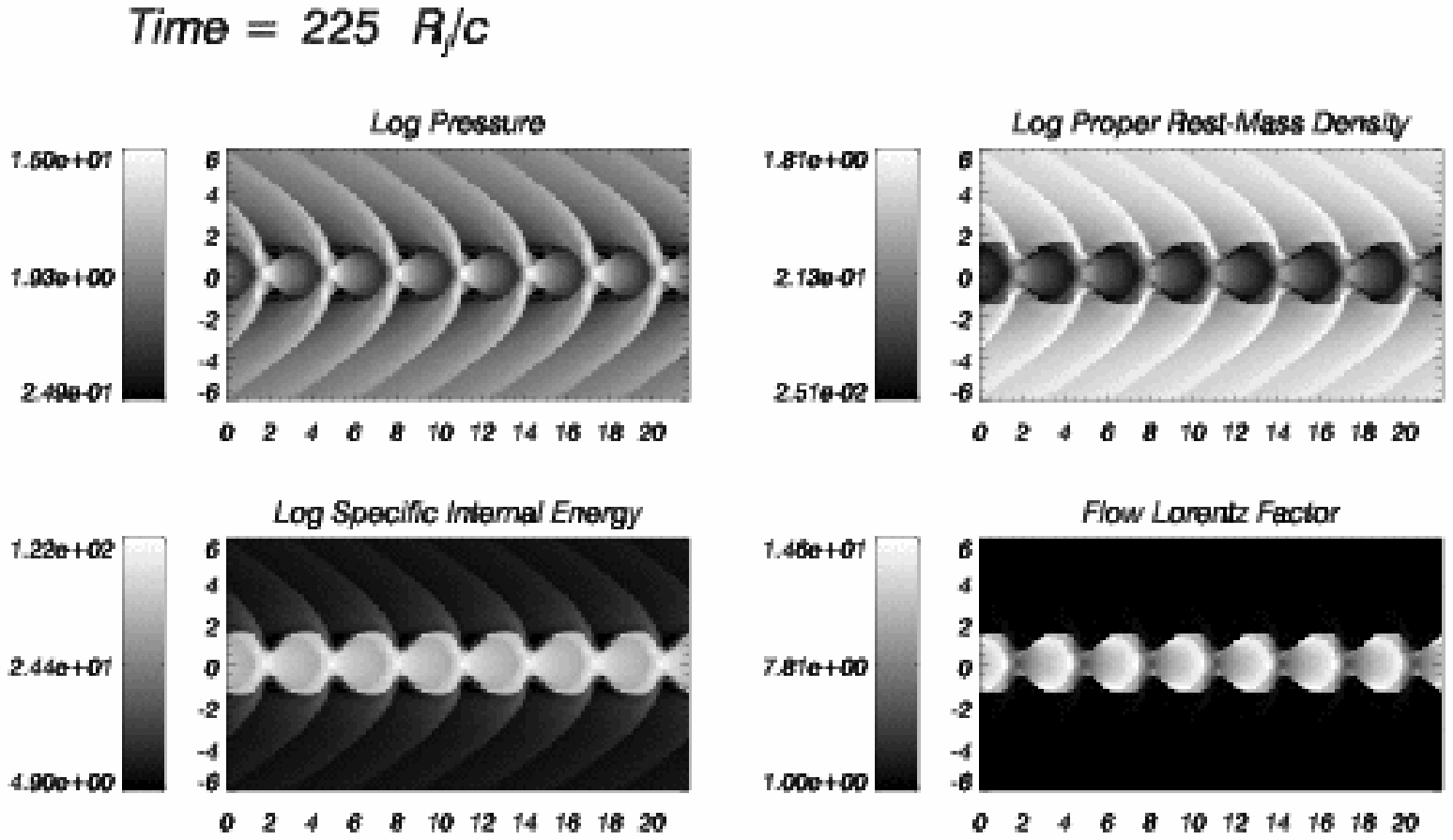,width=0.8\textwidth,angle=0,clip=0}
}
\caption{Same as Fig. (\ref{fig:fab05s}) for models B10 (upper), 
C10 (middle) and D10 (lower).} 
\label{fig:fbc10s}

\end{figure*}

\begin{figure*}
\centerline{
\psfig{file=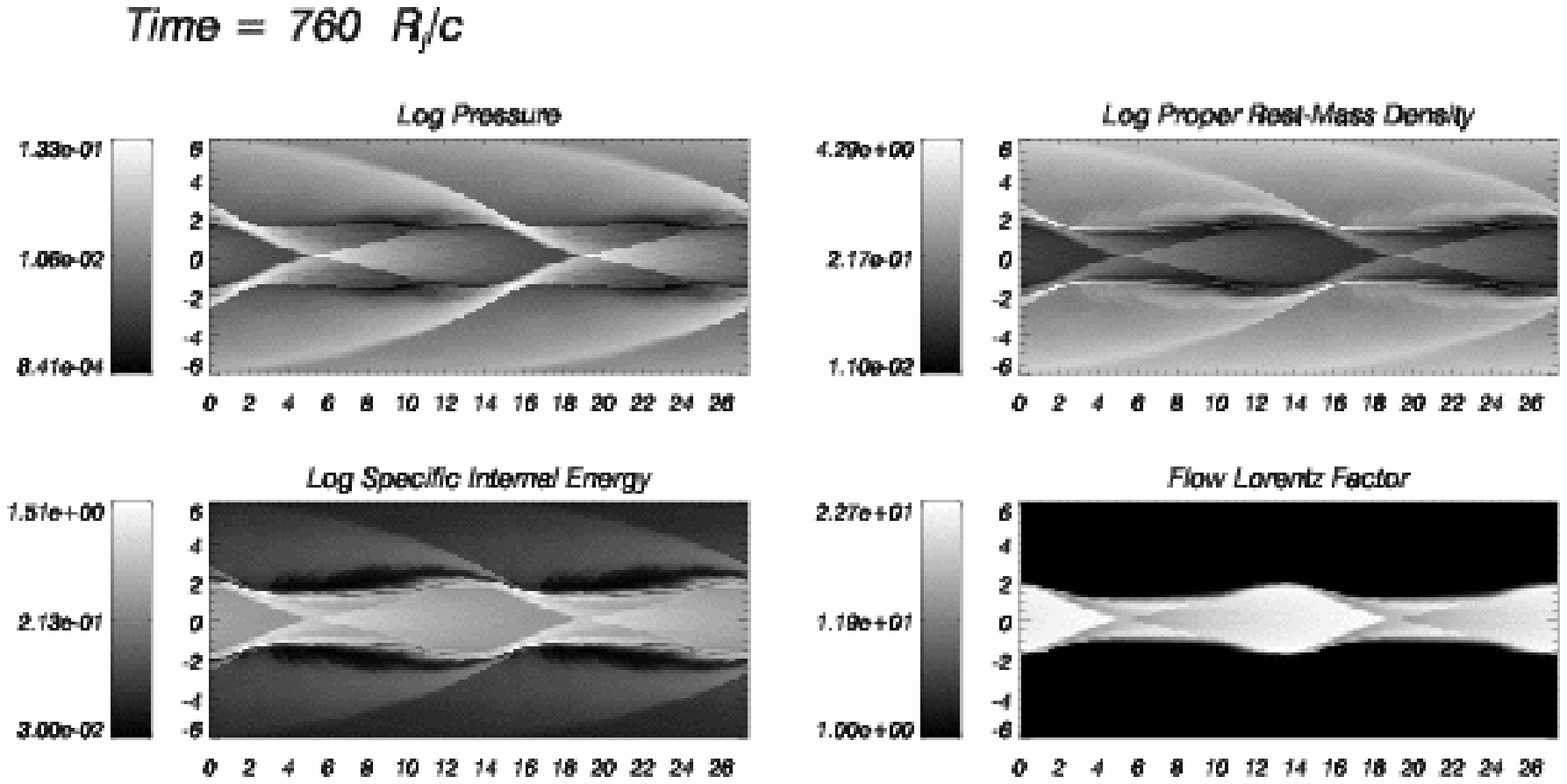,width=0.8\textwidth,angle=0,clip=0} 
}
\centerline{
\psfig{file=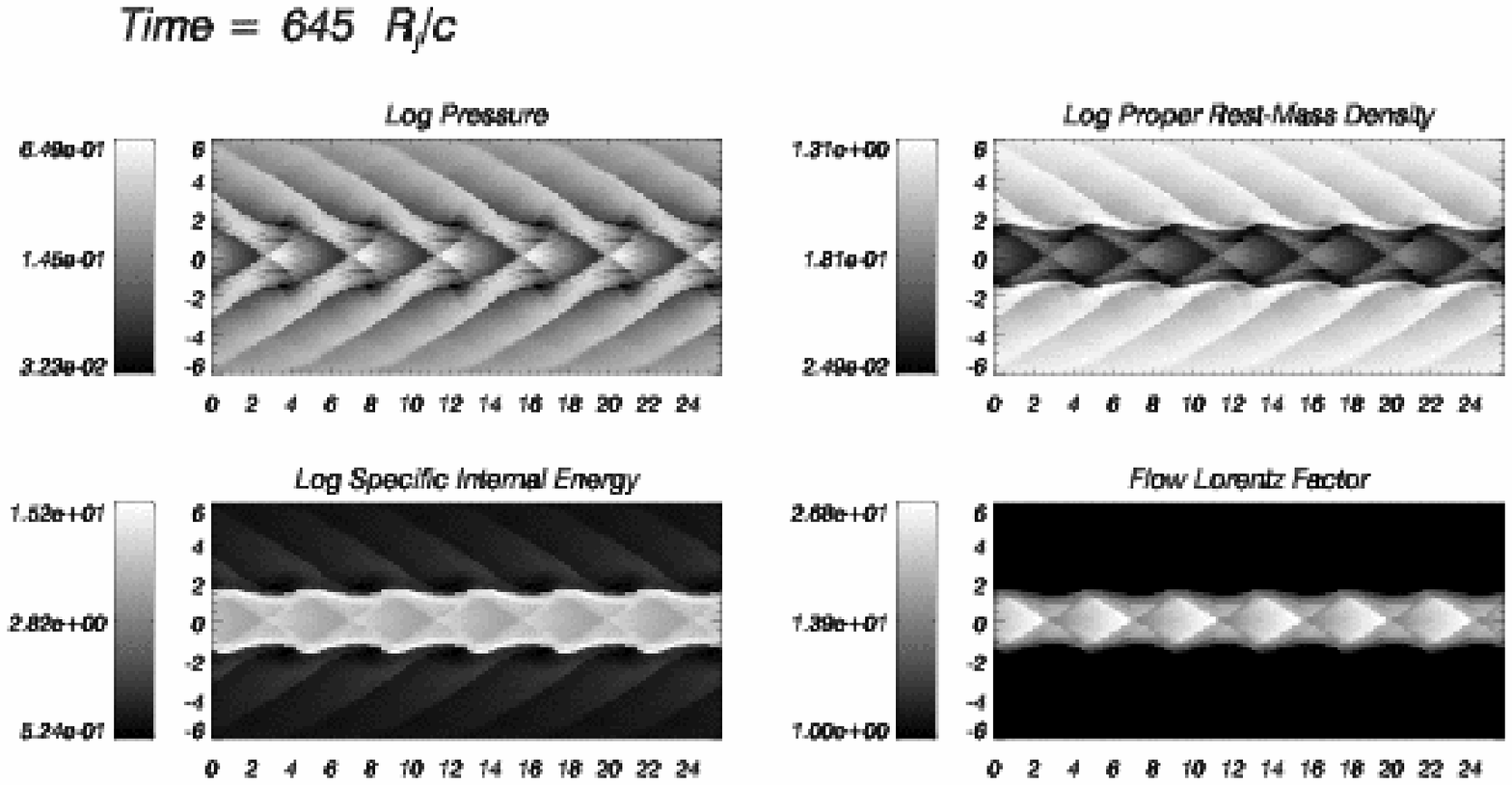,width=0.8\textwidth,angle=0,clip=0} 
}
\centerline{
\psfig{file=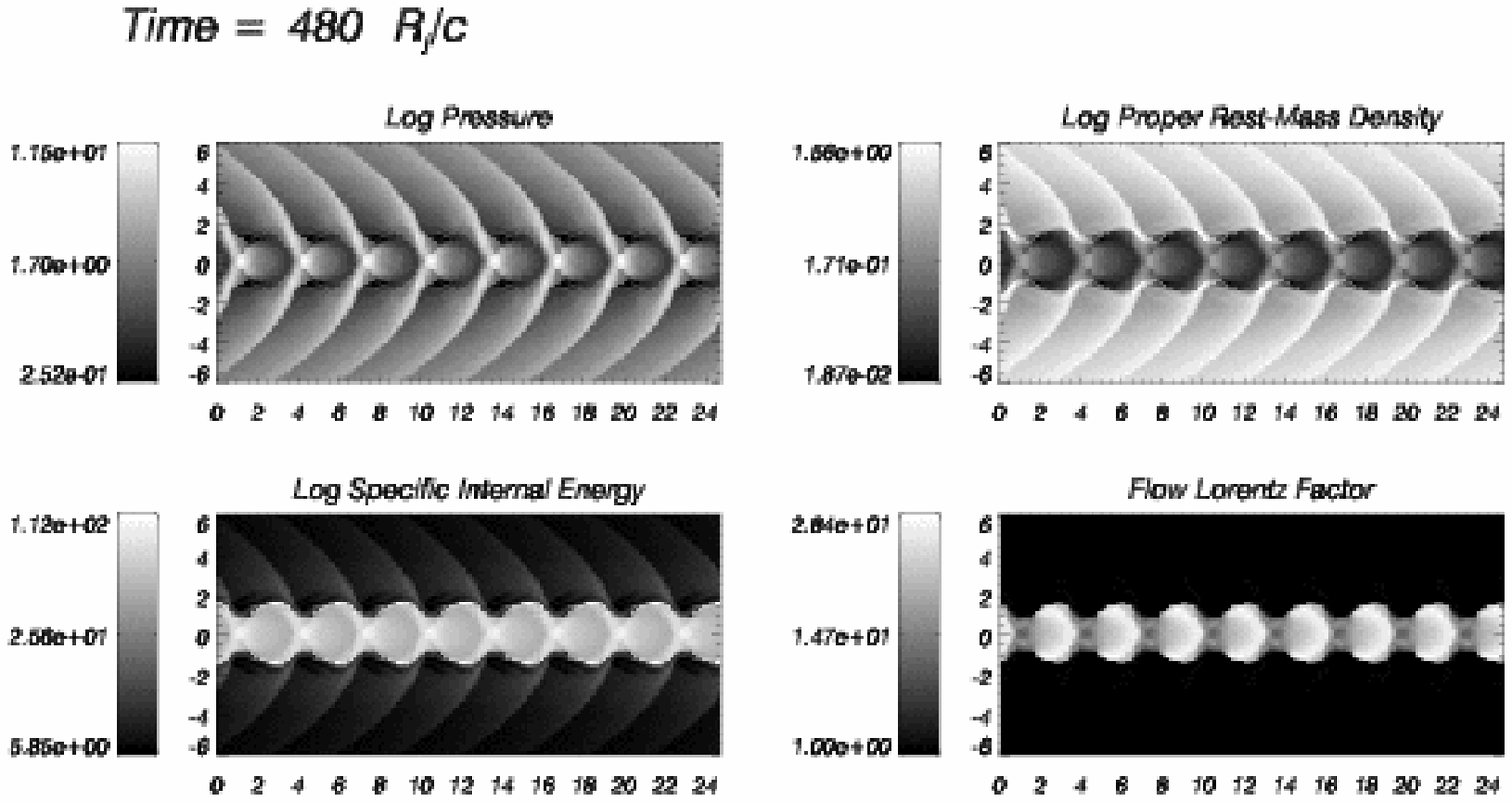,width=0.8\textwidth,angle=0,clip=0}
}
\caption{Same as Fig. (\ref{fig:fab05s}) for models B20 (upper), 
C20 (middle) and D20 (lower).} 
\label{fig:fcd20s}

\end{figure*}

  During the saturation phase, the jet inflates (Bodo et al. 1994
called this phase the expansion phase) and deforms due to transversal
oscillations. On the other hand, the saturation time $t_{\rm sat}$
coincides (within a few time units) with the appearance of an absolute
maximum in the pressure distribution (at $t_{\rm peak}$) at the jet
boundary, and with the start of the mixing phase (to be studied in
Paper II).  Figure~\ref{fig:fab05s}-\ref{fig:fcd20s} display snapshots
of several quantities close to the end of the saturation phase before
the distortion of the jet boundary.

  The structure of perturbations at the end of the saturation phase is
quite similar in all models. The longitudinal
wavelength of perturbations is given by the linear stability analysis
and it is constant because of the fixed length of the computational
domain. Along with the longitudinal wavelength, the opening angle of
oblique waves, far from the jet symmetry plane, is given by the linear
analysis. The opening angle of oblique waves is enlarged in the vicinity
of the jet interface. Closer to the jet symmetry plane the
perturbations are already in the nonlinear phase, which can be
recognized by the apparent presence of oblique shock fronts in
the jet itself and the ambient medium.  These shocks form as a natural
consequence of the nonlinear steepening of oblique sound waves as
their amplitude becomes large. Practically all the four quantities
displayed for a given model (i.e. pressure, rest-mass density, Lorentz
factor and internal energy) show the same structure, so there is no
need to discuss them separately. Therefore we treat the pressure
distribution as a representative quantity. The following properties
of the flow patterns can be noticed in
Figs.~\ref{fig:fab05s}-\ref{fig:fcd20s}:

\begin{enumerate}

\item The interface between jet and ambient medium forms a regular
sinusoidal pattern. The amplitudes are all comparable and have
values of the order of $0.5 - 1.0 R_j$. In all cases the departure
from a regular sinusoidal shape is rather small.

\item The structure of oblique shocks crosses the jet interface in
such a way that the deformed interface is almost parallel to the shock
front. One should notice however that the oblique shock moves in the
flow direction 
in the external medium and in the counter-flow direction in
the jet medium, as can be deduced from the distribution of hot
post-shock and cold pre-shock material. 
These two oppositely moving shock waves fit together because of the
jet background flow.

\item The highest gas pressure is always located on the jet symmetry
plane (the brightest spots on pressure plots), where the crossing
shocks form the familiar x-pattern, but the pressure enhancement on
the jet interface is almost as strong as on the jet symmetry 
plane.

\item The shocks are much stronger (i.e., the jumps between pre- and
post-shock pressure larger) in the cold cases. In the case of hot models
the jumps are smoother. This is a straightforward consequence of the
larger ratios of the perturbated velocity to the sound
speed in the cold cases.

\item There is no significant variation of the structure of nonlinear
perturbations for different values of the Lorentz factor.

\item Model C20 displays a structure of internal waves, which is
absent in other cases. This fine structure is probably connected to
the distinct evolution of this model in the postlinear phase (see
Sect.~\ref{ss:linear} and Fig.~\ref{fig:amplitudes}).


\end{enumerate}

  Therefore the differences between models just before the end of the
saturation phase can be considered as minor.

%

%

\section{Summary \label{s:concl}}

  This is the first of a series of papers devoted to the effects
of relativistic dynamics and thermodynamics in the development of KH
instabilities in planar relativistic jets, accross both linear and
fully nonlinear regimes.  To this aim, we have performed a linear
stability analysis and high-resolution numerical simulations for the
most unstable first reflection modes in the temporal approach, for
three different values of the jet Lorentz factor $\gamma$ (5, 10 and
20) and a few different values of specific internal energy of the jet
matter (from $0.08$ to $60.0 c^2$). In this paper we concentrate in
the early stages of evolution of the KH instability, namely the linear
and saturation phases. The present paper is also intended to set the
theoretical and numerical background for the whole series of papers.

  Our simulations describe the linear regime of evolution of the
excited eigenmodes of the different models with high accuracy. The
growth rates of the perturbed modes in the vortex sheet approximation
were determined with an average relative error of 20\%.

  In all the examined cases the longitudinal velocity perturbation is
the first quantity that departs from linear growth when it reaches
a value close to the speed of light in the jet reference frame. The
reason for this saturation, specific to relativistic dynamics, is not
so obvious in the reference frame of the external medium where it
saturates at a smaller value ($\approx c/\gamma$, where $\gamma$ is
the bulk flow Lorentz factor in the jet).

  The saturation phase extends from the end of the linear phase up to
the saturation of the transversal velocity perturbation (at
approximately $0.5c$ in the jet reference frame). The saturation times
for the different numerical models have been explained from elementary
considerations, i.e. from properties of linear modes provided by the
linear stability analysis and from the limitation of the transversal
perturbation velocity.

  The limitation of the components of the velocity
perturbation at the end of the linear and saturation phases allows us
to conclude that the relativistic nature of the flow appears to be
responsible for the departure of the system from linear evolution.  
This behaviour is consistent with the one deduced by Hanasz (1995,
1997) with the aid of analytical methods.

  At the saturation time the perturbation structure is close to the
structure of the initial perturbation (the one corresponding to the
most unstable first reflection mode), except that the oblique sound
waves forming the perturbation became steep due to their large
amplitude. It is interesting to note that the oblique shocks are
stronger (i.e., the pressure jumps are larger) in the colder cases.

  Our simulations, performed for the most unstable first reflection
modes, confirm the general trends resulting from the linear stability
analysis: the faster (larger Lorentz factor) and colder jets have
smaller growth rates.  As we mentioned in the Introduction, Hardee et
al. (1998) and Rosen et al. (1999) note an exception which occurs for
the hottest jets. These jets appear to be the most stable in their
simulations (see also the simulations in Mart\'{\i} et al. 1997). They
suggest that this behaviour is caused by the lack of appropriate
perturbations to couple to the unstable modes. This could be partially
true as fast, hot jets do not generate overpressured coocons that let
the jet run directly into the nonlinear regime. However, from the
point of view of our results, the high stability of hot jets may have
been caused by the lack of radial resolution that leads to a damping
in the perturbation growth rates.  We note that the agreement between
the linear stability analysis and numerical simulations of KH
instability in the linear range has been achieved for a very high
radial resolution of 400 zones/$R_j$, which appears to be especially
relevant for hot jets. Finally, one should keep in mind that the
simulations performed in the aforementioned papers only covered
about one hundred time units, well inside the linear regime of the
corresponding models for small perturbations. 

  The high accuracy of our simulations in describing the early stages
of evolution of the KH instability (as derived from the agreement
between the computed and expected linear growth rates and the
consistency of the saturation times) establishes a solid basis to study
the fully nonlinear regime, to be done in the second paper of this
series. In this paper (Paper II) we will show that the similarities
found in the evolution of all the models accross the linear and
saturation phases is lost and very different nonlinear evolution is
found depending on the initial jet parameters.

\begin{acknowledgements}
  This work was supported in part by the Spanish Direcci\'on General
de Ense\~nanza Superior under grant AYA-2001-3490-C02 and by
the Polish Committee for Scientific Research (KBN) under grant PB
404/P03/2001/20. M.H. acknowledges financial support from the visitor
program of the Universidad de Valencia. M.P. has benefited from a
predoctoral fellowship of the Universidad de Valencia ({\it V Segles}
program). The authors want also to acknowledge financial support from
the Spanish-French {\it Picasso} program and the Polish-French {\it
Joumelage} program in the period 1997-98 when the research on
relativistic jet stability was initiated.
\end{acknowledgements}

\begin{appendix}
\section{Influence of the numerical resolution in the
          description of the linear regime of KH modes of 
          relativistic flows}

\begin{figure*}
\centerline{
\psfig{file=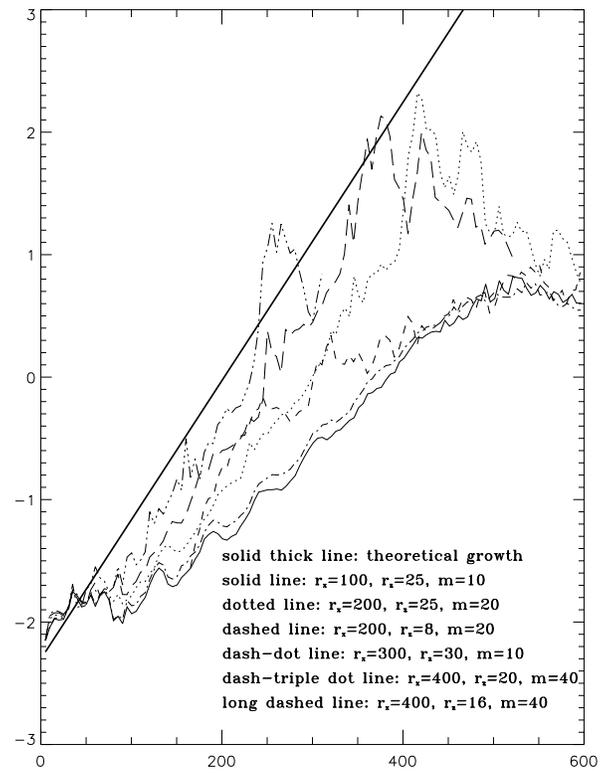,width=\columnwidth,angle=0,clip=0}
}
\caption{Linear growth of the amplitude of the pressure perturbation
  (in logarithmic scale) versus time (in units of $R_j/c$) for
  different resolutions. $r_x$ stands for transversal resolution,
  $r_z$ for longitudinal resolution, and $m$ is the value which gives
  the shear layer steepness.}
\label{fig:linear2}
\end{figure*}

  Growth of the instability depends critically on the numerical
viscosity of the algorithm. Hence our first aim was to look for
suitable numerical resolutions by comparing numerical and analytical
results for the linear regime. We performed a number of simulations
based on model A05 changing longitudinal and radial resolution, and
also the exponent $m$ for the shear layer steepness (see
Eqn. \ref{shearlayer1} and \ref{shearlayer2}). Figure
\ref{fig:linear2} shows results for several of those simulations.

  A shear layer was included in order to avoid non-steadiness in initial
conditions given by discontinuous separation between the beam and the
external medium. Therefore, if for some transversal resolution we
introduced a shear layer which was too thin, we found a similar
non-steady behavior. However, in order to reproduce the linear regime,
we needed a steep enough shear layer, due to the fact that theory was
developed for a discontinuous separation between both media. We can
see in figure \ref{fig:linear2} that $m=10$ models are considerably
damped with respect to the theoretical growth, independent of the
transversal resolution. Hence resolution perpendicular to the flow
appeared to be essential, requiring very high resolutions (400
zones/$R_j$) and thin shear layers ($m=40$) with 40 to 45 zones, i.e.,
an equilibrium between steepness and number of cells. Lower
transversal resolutions and/or thicker shear layers led to
non-satisfactory results, with a slow or damped growth. A very low
longitudinal resolution results in damping of growth rate, too, as can
be seen from comparison of $m=20$ models, but $r_z=16$ seems to give a
reasonable growth rate compared to theory. This (small) resolution of
16 zones/$R_j$ along the jet was taken as a compromise between
accuracy and computational efficiency.

%
%

\end{appendix}



\end{document}